\documentclass[twocolumn,prb]{revtex4}

\usepackage[dvips]{graphicx}

\newcommand{\PRL}[1]{Phys. Rev. Lett. {\bf #1}}
\newcommand{\PRB}[1]{Phys. Rev. B {\bf #1}}
\newcommand{\PRE}[1]{Phys. Rev. E {\bf #1}}
\newcommand{\eqref}[1]{Eq.\ (\ref{#1})}
\newcommand{\figref}[1]{Fig.\ \ref{#1}}
\newcommand{\secref}[1]{Sec.\ \ref{#1}}
\newcommand{\refref}[1]{Ref.\onlinecite{#1}}
\newcommand{\be}{\begin{equation}}
\newcommand{\ee}{\end{equation}}
\newcommand{\ba}{\begin{eqnarray}}
\newcommand{\ea}{\end{eqnarray}}
\newcommand{\pr}{{\rm Prob}}

\begin{document}

\title{	
The three-dimensional random field Ising magnet:
interfaces, scaling, and the nature of states
}

\date{\today}

\author{A. Alan Middleton}
\affiliation{Department of Physics,
Syracuse University, Syracuse, New York 13244}
\author{Daniel S. Fisher}
\affiliation{Lyman Laboratory of Physics, Harvard University,
Cambridge, Massachusetts 02138}

\begin{abstract}

The nature of the zero temperature ordering transition in the
three dimensional Gaussian random field Ising magnet is studied numerically,
aided by scaling analyses.
Various numerical calculations are used to consistently
infer the location of the transition to a high precision.  A variety
of boundary conditions are imposed on large samples to study the order
of the transition and the number of states in the large volume limit.
In the ferromagnetic phase, where the domain walls have
fractal dimension $d_s = 2$, the scaling of the
roughness of the domain walls, $w\sim L^\zeta$,
is  consistent with the theoretical prediction $\zeta = 2/3$.
As the randomness is increased
through the transition,
the probability distribution of the interfacial tension of domain walls
scales in a manner that is clearly
consistent with a single second order transition.
At the critical point, the fractal dimensions of domain walls
and the fractal dimension of the outer surface
of spin clusters are investigated: there are at least two distinct physically
important fractal dimensions that describe domain walls.
These dimensions are argued to be related by scaling to combinations of the energy scaling 
 exponent, $\theta$, which determines the violation of hyperscaling, the correlation length
exponent $\nu$, and the magnetization exponent $\beta$.
The value $\beta = 0.017\pm 0.005$ computed from finite
size scaling of the magnetization is very nearly
zero: this estimate is supported by the study of the
spin cluster size distribution at criticality.
The variation of
configurations in the interior of a sample with boundary conditions is
consistent with the hypothesis that there is a single transition
separating the disordered phase with one ground state from the ordered
phase with two ground states.
The array of results, including values for several
exponents, are shown to be consistent with a scaling picture
and a geometric description of the influence of boundary
conditions on the spins.  
The details of the algorithm used
and its implementation are also described.

\end{abstract}


\maketitle

\section{Introduction}
In spite of many years of study, the behavior of phases and phase transitions
that are dominated by quenched randomness is still controversial.  One such
lively controversy has concerned the existence
or lack thereof of an ordered phase
in the random field Ising model (RFIM) in three dimensions.
Although this was
eventually resolved in the affirmative by rigorous
work, \cite{proofs} the nature of the
phase transition and the possibility of a phase intermediate between the
paramagnet and the ferromagnet is still controversial.
 
Numerical simulations of the random field Ising model --- and experiments ---
are impeded by the dramatic slowing down that occurs as the phase transition
is approached due to the existence of free energy barriers which are broadly
distributed but typically grow as a power of the correlation length.  Such
barriers are general characteristics of phases controlled, in a
renormalization group (RG) sense, by stable zero temperature fixed points. For
the random field Ising model, not only is the low temperature phase
controlled  by a
zero-temperature fixed point (as is the case for conventional
pure systems), but the phase transition itself is also
controlled by such a fixed point. \cite{FisherRFIM,Villain} Indeed, the
ground state properties of the
random field Ising model undergo a phase transition as the strength of the
randomness is increased and it is this  zero-temperature transition that
governs the behavior of the transition at positive temperatures.
Fortunately, this means that much can be learned by studying the ground state
properties.
 
It has been known for some time that combinatorial algorithms can be used
effectively to find the ground states of various classes of random systems and
the RFIM was one of the first to be studied in this way. \cite{Ogielski} With current
computers, the algorithm is very fast and large system sizes can be studied
in enough detail to obtain good statistics enabling the tools of finite size
scaling to be used to analyze the zero-temperature phase transition.
 
Various significant open questions exist about the phase transition in the
RFIM.  Although a self-consistent scaling picture of a 
zero-temperature critical fixed point was proposed early on, it has not been
adequately tested and other scenarios have been suggested, including
a first order phase transition \cite{dAuriacSourlas,Sourlas}
and an intermediate phase with ``replica-symmetry
breaking" \cite{MezardYoung,MezardMonasson} presumably meaning
many coexisting equilibrium states.
 
In this paper we study the RFIM with Gaussian distributed random
fields, focusing on the nature of the phase transition and the
sensitivity of the ground states to varying boundary conditions (BCs) as a
probe of the number and nature of the infinite system states. As will
be explained in some detail, our results strongly support the scaling
picture of the transition. In this picture,
there is a single second order critical point characterized
by three scaling exponents: $\nu$ for deviations from the critical
point, $\theta$ for the energy at the critical point, and $\beta/\nu$
for the magnetization at the critical point.  We clarify some of the
substantial confusion about the order of the transition by showing
that both $\theta$ and $\nu$ as well as the distributions of the
``stiffness" and spin clusters are very different from what they
would be at a first order transition.  Nevertheless, as observed
previously, the magnetization exponent $\beta$ is extremely small so
that even with the very large sizes we study, the magnetization
appears almost discontinuous.

\section{Model and Numerical Method}

The random field Ising model\cite{NattermannInYoung}
has Hamiltonian,
defined over spin configurations $\left\{s_i=\pm 1\right\}$,
\begin{equation}
H = - J\sum_{\left<ij\right>} s_i s_j - \sum_{i}h_i s_i,
\end{equation}
with the random fields $h_i$
chosen independently from a distribution which we take to be
Gaussian with mean zero and variance $h^2$.
The ferromagnetic exchange coupling, $J$, is fixed at unity
in the simulations and
the sites $i$ lie on a cubic
lattice with interactions between nearest neighbor
pairs $\left\{\left<ij\right>\right\}$.
The basic nature of the phase diagram of the 3-D RFIM
is well-known: As the temperature is lowered for small $h$,
there will be a critical temperature, $T_c(h)$,
below which the RFIM becomes ferromagnetically ordered with a non-zero spontaneous magnetization.
As the strength of the random field increases, the critical temperature decreases until at a critical field, $h_c$, it goes to zero. 

Both the paramagnetic and the ferromagnetic phases have been proven to
exist at both zero and positive temperatures\cite{proofs} and the
transition between them can thus be studied by varying $h$ at $T=0$.
The simplest scenario at zero-temperature is a single critical field strength $h_c$ above
which the spins are disordered with a unique infinite-system ground
state and exponential decay of correlations, and below which there are
two infinite-system ground states, one with predominantly up spins and
the other with predominantly down spins.

The nature of the phases and the phase transition(s) between them can
be probed by studying the effects of various boundary conditions on
larger and larger systems --- most simply cubes of size $ {\cal
V}=L\times L\times L$. In the disordered phase the orientation of a
spin far from the boundaries is typically determined by the collection
of random fields within a correlation length $\xi(h)$ of the spin and
is insensitive to boundary conditions imposed far away.  In contrast,
in the ferromagnetic phase some spins will still be controlled by the
random fields in their vicinity, but a finite fraction of the spins
will be controlled by the boundary conditions --- no matter how far
away they are imposed.  The simplest scenario is a single transition
between these two phases.  The primary goal of this paper is to
examine in detail the nature of this zero-temperature phase
transition.

Many previous studies of the ground states of the RFIM (as well as
finite temperature studies) have focused on the magnetization per spin,
$m={\cal V}^{-1}\sum_i s_i$, and the results have been somewhat
ambiguous. Some have interpreted the numerical results as indicating a
second order transition,\cite{Ogielski,Swiftetal,HartmannNowak} while
others have concluded that the transition is first
order.\cite{Sourlas}  Some Monte Carlo results\cite{RiegerRFIM}
suggest that the finite-temperature transition is second order, but
with the magnetization exponent $\beta$ nearly zero.
Others,\cite{MachtaNewmanChayes} using a varying external field, have
found a coexistence of states suggestive of a first order transition.
It is clear from these studies that if the transition is second order,
the order parameter exponent $\beta$, $\left|m\right| \sim
(h_c-h)^{\beta}$ must be very small, making definitive conclusions
based on magnetization alone difficult.  We have thus focused much of
our attention --- particularly for locating the transition and finding
the exponents --- on other properties which naturally distinguish the
phases.

\subsection{Algorithm}

The (almost surely) unique ground state of a finite sample
can be determined in time polynomial in the number of spins.\cite{dAuriac}
The method is based on a reduction of the problem of determining RFIM
ground states to a maximum-flow problem on an augmented graph.
One can then use combinatorial optimization algorithms
\cite{Goldberg,combopt,AAMrough,RiegerReview}
to solve the maximum-flow problem.
We describe the special features of the
algorithm implementation, its verification, sample timings,
and the use of integer valued $h_i$ in the Appendix.

\subsection{Statistics and analysis}

We have studied system sizes
up to $256^3$, which contain over $1.6 \times 10^7$ spins.
Independent samples were simulated for each value of $h$. Separate
realizations were also generated for boundary induced domain
walls, spin
cluster properties, magnetization, and the thermodynamic limit studies.
The same samples and domain walls were used in the stiffness and domain
wall property studies. For smaller systems ($8^3$ through $32^3$),
$10^5$ samples were optimized, typically. (For the domain roughness
measurements in the ordered phase, $10^2$ or $10^3$ samples provided
sufficient data, as fluctuations in the interface width are not
large.) Of order $10^3$ to $10^4$ samples
were studied for each quantity for the $64^3$ and $128^3$ samples.
For $L=256$, $4 \times 10^2$ to $10^3$ samples were studied at each $h$,
as part of the magnetization and cluster studies.

Error bars for exponent
values throughout this paper
include both estimated systematic errors due to apparent
finite size effects and errors due to statistical uncertainties;
the error bars represent an estimated range of values in which the
value lies, with high confidence. In contrast, error bars in the figures
reflect $1\sigma$ statistical
uncertainties, which we find to be generally consistent with
confidence intervals found by resampling.

Generally (except for the stiffness, the roughness in the
ferromagnetic phase, fitting a power law to the bond energy
density, and the $P_{D/O,\pm}$ plots),
we have used estimates of effective exponents as a function of
system size to estimate exponents, rather than scaling plots. This is
done to more clearly see trends in the data that reflect finite size
corrections. Finite size corrections tend to be monotonic and introduce
a drift with $L$ in the effective exponents. 
Given the good statistics of the data sets that can be generated with
optimization algorithms,
collapsing data can obscure these corrections, as the drift can be
corrected with a slightly erroneous exponent.
Where we have used scaling plots, we do not try to collapse all of the
data onto a single curve, but keep in mind that the finite size corrections
give a consistent drift with system size and
we therefore tried to optimize the fit to the largest
systems and near $h_c$.

\section{Summary of results}

As we are interested in the behavior of the RFIM in the thermodynamic
limit, we have studied the approach to the infinite-volume limit using
finite-size scaling analysis techniques similar to those applied to
spin glasses and other random
systems.\cite{McMillanDWRG,BrayMooreLCRG,BrayMoorestates,AAMstates,PalassiniYoung2d,PalassiniYoung3dPRL}
However, in
contrast with ground state studies of spin glasses and other random
models for which only a single thermodynamic phase exists, the results
presented here give insight into the transition between
two phases.\cite{2DFMSG}

\subsection{Stiffness}

The fundamental difference between an ordered phase and a disordered
one is the {\it stiffness} (or rigidity) of the former: the free energy cost of
changing one part of a system with respect to another part far away.
At a macroscopic level, this free energy cost must be at least of
order $k_B T$ and is usually much larger, diverging as a power of the
system size.  For an Ising ferromagnet, this stiffness is provided by
the free energy cost of a domain wall which scales as its surface
area. Thus a natural quantity to study for the ground states of the
RFIM is the domain wall energy. This can be obtained from the
difference in energy between antiparallel and parallel boundary
conditions imposed on opposite sides of a system of cross sectional
area $L^2$.
A particular combination of these we call the {\em stiffness}, which
we denote by $\Sigma$.
Because of the randomness, this energy
will be sample dependent and there is information to be
gleaned from its distribution as well as its mean.

The scaling theory of the putative critical point of the random field
Ising model predicts that the distribution of the stiffness
will have a scaling form near the critical point:
\be
\label{sigma-scaling}
\pr[d\Sigma]\approx
\frac{d\Sigma}{CL^\theta}P\biggl(\frac{\Sigma}{CL^\theta},K(h-h_c)L^{1/\nu}
\biggr)
\ee
with $\theta$ and $\nu$ universal exponents, $P$ a
universal scaling function (which does, however, depend on the shape of the sample), and $C$ and $K$ non-universal
coefficients.  In the {\em ferromagnetic phase}, the distribution of the
stiffness will be sharply peaked at long length scales about a mean
value which grows as $\sigma (h)L^2$ with $\sigma(h)$ the
interfacial tension.
This interfacial tension vanishes as $h\nearrow h_c$.
In the {\em disordered phase},
$\Sigma$ will typically fall off exponentially for system thicknesses,
$L$, much larger than the correlation length $\xi(h) \sim
(h-h_c)^{-\nu}$.
This  exponential decay of the stiffness with $L$
is confirmed for all values $h>h_c$ examined in our numerical results.

At the {\em critical point}, the
distribution of $\Sigma$ will be broad with both mean and width of order
$L^\theta$.  The exponent $\theta$ thus characterizes the scaling of
the stiffness at the critical point.  As long as $\theta$ is positive,
the basic features of the zero-temperature critical point will be
stable to thermal fluctuations and the finite-temperature transition
will be in the same universality class. \cite{FisherRFIM}

Our studies of the stiffness are based on computing energies for samples
periodic in two directions and having fixed uniform boundary spins on
the other two faces.
Our results
are very consistent
with the scaling predictions for the larger system sizes, up to $128^3$,
close to the critical point, which occurs at
\be\label{hc} h_c \simeq 2.270 \pm 0.004.\ee
This location for the critical point is consistent with those obtained from scaling analyses of
the domain wall dimension and the magnetization.
It is somewhat lower than some previously
reported estimates such as  $h_c\approx 2.33$,\cite{Ogielski,Swiftetal}
but it is consistent with the values $h_c = 2.29 \pm 0.04$ reported by
Hartmann and Nowak\cite{HartmannNowak}, $h_c=2.26\pm0.01$ reported
by d'Auriac and Sourlas,\cite{dAuriacSourlas} and $h_c = 2.28 \pm 0.01$
reported by Hartmann and Young.\cite{HartmannYoung}
Taking $h_c=2.270$, the exponents that give a good
scaling fit for the stiffness are found to be
\be \theta \simeq 1.49 \pm 0.03 \ee
and
\be \nu \simeq 1.37 \pm 0.09.\ee
The value for $\theta$ is consistent with
exact bounds as well as with values derived from finite temperature
simulations 
by applying exponent relations to
measured critical behavior.\cite{RiegerRFIM}
Note that if the
transition had been first order, one would have expected to find
$\theta=d-1=2$, with a double peaked distribution of $\Sigma$
corresponding to ``ordered" and ``disordered" samples, and an effective
$\nu=2/d=2/3$; the results we find are far from these.

From the modified hyperscaling law
appropriate to transitions governed by
zero temperature fixed points,\cite{FisherRFIM,Villain}
\be(d-\theta)\nu = 2-\alpha\label{hyper}\ee
with $d$ the dimension, here equal
to three, we predict that the specific heat exponent for the
finite-temperature transition (and for the second derivative of the
energy with respect to $h$ at zero-temperature near the transition) is
\be\alpha = 2-(3-\theta)\nu \simeq -0.07 \pm 0.17.\ee
We also fit the 
sample averaged
bond part of the energy, $\overline{E}_J$, at $h_c$ to the form
$\overline{E}_J \sim c_1 - c_2 L^{(\alpha-1)/\nu}$ to
more directly obtain $\alpha$, inspired by the recent approach
of Hartmann and Young,\cite{HartmannYoung} who examined the scaling
of the derivative $dE_J/dh$.
We find a consistent value for $\alpha$ using similar methods, although
our value disagrees substantially with that of Hartmann and Young.
We also use an extrapolation of $\overline{E}_J$,
based on the dimension
of the domain wall surfaces,
which define $E_J$, to find
\be
\alpha = -0.01 \pm 0.09.
\ee
Both values are consistent with experiments,\cite{JaccarinoBiref}
which yield a small value of $\alpha$.

\subsection{Domain walls}

In addition to the stiffness measurements, we have investigated the
properties of the domain walls that are forced by appropriate changes
of the boundary conditions.  In the ferromagnetic phase, we expect
that these will be flat on large length scales and have area
proportional to $L^2$. These walls will be rough
with a transverse width $W$ on a scale $\ell$ described by the roughness
exponent $\zeta$, $W\sim \ell^\zeta$.
But at the critical point, we expect the walls
will become fractal.  The definition of a domain wall in the RFIM has
ambiguities because some isolated clusters of spins --- in particular
those with anomalously strong random fields --- are {\it frozen},
i.e., unaffected by changes in boundary conditions.  The identification
of the bonds that define the domain wall is therefore uncertain up to
these fixed spins.  We use three methods for calculating the fractal
dimension of the domain walls introduced by changes in the boundary
conditions; each definition has a distinct physical import in a scaling
picture of the RFIM.

One method is to determine the surface area of the set of spins
connected to one face of the sample that are {\it unchanged} when the
spins on the {\em opposite face} are reversed. This yields a {\it spanning
surface} of a dimension that we denote $d_s$.  The second method is
based on a box counting approach that counts which volumes in a system
with antiparallel boundary conditions differ from {\it both} the ``up''
and the ``down'' configurations obtained from parallel spin boundary
conditions; this we denote $d_I$, to indicate its role as a measure
of the volume locally incongruent with these two configurations.

A third method does not measure a
dimension directly but rather an energy: the contribution of the
exchange interactions to the stiffness at the critical point.  As will
be explained later, this fractal dimension, $d_J$, is {\it not}
expected to be an independent exponent.  Rather, it is related to the
others by
\be
d_J=\theta + 1/\nu,
\ee
a relation obtained by
considering the derivative of the stiffness with respect to $h$.  It
can be seen that our results for the exponents are entirely consistent
with this scaling law.

The three exponents associated with the fractal
dimension of the critical domain walls are similar, but perhaps 
not all mutually consistent, given the estimated error bars:
\begin{eqnarray}
d_s = 2.30 \pm 0.04 \\
d_I = 2.24 \pm 0.03 \\
d_J = 2.18 \pm 0.03  
\end{eqnarray} 
As we will discuss, we believe that at least two
of these dimensions indeed measure slightly
distinct quantities, due to frozen spin clusters that
are relatively independent of boundary conditions.
The simplest plausible conjecture is that $d_J=d_I < d_s$, though
it may be that $d_J$ and $d_I$ are distinct.

\subsection{Magnetization}

As mentioned above, some previous studies have found that the magnetization
appears to be discontinuous at the transition.  Indeed, our data for
the magnetization as a function of size for various types of boundary
conditions that were chosen so that some favor a ferromagnetic state
while others favor a disordered state, {\it appear} to be consistent with
the coexistence of three states at the critical point as was found in
other recent work.\cite{MachtaNewmanChayes} But, as discussed below, we believe
that this conclusion is influenced by the nearness of $\beta/\nu$ to zero.
Based upon the  scaling picture
and numerical evidence, we will argue 
that, at the critical point there is only a single state and the apparent ``up'' and ``down'' configurations do
{\it not} correspond to distinct states in the infinite volume limit.

Using the magnetization data and the best fit critical point found
from our studies of the stiffness, we can attempt to extract an
estimate for the scaling of the magnetization with system size at the
critical point.  This yields
\be
\frac{\beta}{\nu} = 0.012 \pm 0.004\ \ \ \ \ {\rm[magnetization]}, \label{beta/nu}
\ee
which is {\em inconsistent with zero} at the level of three standard deviations.
The primary uncertainty in our estimate of $\beta/\nu$ arises from the
uncertainty in the value of
$h_c$, as the statistical errors in the sample average $\overline{|m|}$ are
relatively small at fixed $h$. This exponent describes the magnetization very
well for systems of size $32 \le L \le 256$, for a range of $h_c$, 
$2.265 < h_c < 2.275$.

\subsection{Spin clusters and walls}

In spite of the smallness of $\beta$, useful information on the decay
of spin correlations at the critical point can be obtained indirectly
by studying the statistical properties of the domain
walls separating connected clusters of
parallel spins.  In the ferromagnetic phase, we expect that the
probability of finding a region of diameter $\ell$ that is not
affected by the boundary conditions decays exponentially for $\ell
\gg \xi$ as $\exp(-C (\ell/\xi)^{d-2})$.

Since the system {\it appears} to be ferromagnetic at the critical
point, due to $\overline{|m|}$ being nearly unity for the system
sizes studied, we also study clusters of the minority spins at $h\approx
h_c$.
The clusters are defined hierarchically starting from the largest
connected cluster of connected spins,
with the
surface of each cluster given by its {\em outermost} surface, that is, the
set of bonds connecting it to the surrounding cluster.
The volume of each cluster {\it includes} that of
the fully enclosed subclusters of the opposite sign (and their subclusters, if
any, etc). But the outer surface of a cluster does {\it not} include the
{\em surfaces} of its fully enclosed subclusters,
whose number scales with the volume of the cluster. 

The outer cluster surfaces are found to be fractal with mean area $\overline{a}$
(averaged over clusters and samples)
scaling with enclosed volume $v$ as
\be
\overline{a}(v)\sim v^{d_s^c/d}
\ee
with the exponent
\be
\frac{d_s^c}{d} \simeq 0.755 \pm 0.008,
\ee
suggesting a surface fractal dimension $d_s^c \simeq 2.26 \pm 0.02$,
a value consistent with the domain wall dimension $d_s$.  

Perhaps more interesting is the distribution of the {\it number density} of spin
clusters as a function of their size, in particular the probability
$\rho(v)$ that a given site is in a minority spin cluster of size of
order $v$ --- more precisely,
\be
\rho(v)\equiv \frac{v}{\delta v}\pr[{\rm site} \in {\rm cluster \ of\ size \ in}\ (v,v+\delta v)] .
\ee
We find that over the range of sizes studied $\rho(v)$ appears to
converge to a {\it small constant} value, $\rho_\infty$, for $1\ll v
\ll L^3$, with periodic boundary conditions.
This implies that in the limit of an extremely large
system, any given spin will definitely be in such a ``minority spin"
cluster; indeed, it will typically be within one such large cluster
which itself will be within a cluster of typical size $\sim \rho_\infty^{-1}$
larger which
itself will be in an even larger cluster, etc.  This is
exactly the type of behavior that gives rise to power law decay of
spin correlations at a critical point on sufficiently long scales, as is explained in Section IX. It
is consistent with expectations from other observations we have made,
in particular that the probability that the stiffness of a finite sample is
exactly zero tends to a non-zero constant for large system sizes at
the critical point.

The value of
\be
\rho_\infty \simeq 0.0019 \pm 0.0004
\ee
that we find\cite{fit_rho_note} yields an estimate for
\be
\beta/\nu = 2d\rho_\infty \simeq 0.011 \pm 0.003\ \ \ \ \ {\rm[cluster]}
\ee
consistent with \eqref{beta/nu}.
This exponent controls the decay of the typical magnetization with
system size at the critical point:
\be
m(h_c) \sim L^{-\beta/\nu}.
\ee
For $L=128$, this only gives a reduction factor of $0.94$ from the
magnetization of a small system and is consistent with our
magnetization data.  Note that with this estimate, one would need to
go to system sizes of order
$10^{21} \stackrel{\textstyle<}{\sim} L \stackrel{\textstyle<}{\sim} 10^{38}$
to see a factor of two
reduction in the magnetization at the critical point!

For the magnetization in the ferromagnetic phase,
using this calculation of $\beta/\nu$ and consistent with the
finite size scaling of the magnetization,
we expect conventional behavior with
\be
m \sim (h_c-h)^\beta
\ee
but with
\be
\beta \simeq 0.017 \pm 0.005
\ee
--- far smaller than for any other known system with the exception of
the one-dimensional Ising model with long-range $1/r^2$
interactions which has a critical transition with a discontinuous
magnetization, i.e. $\beta=0$.\cite{inverse-square-Ising}
This small value is near the value $\beta=0.02$
suggested from numerical renormalization
group calculations on a hierarchical lattice.\cite{CaoMachta}
The numerical value we find is consistent with several
previous studies: for example,
Hartmann and Nowak determine $\beta=0.02\pm0.01$, using
exact ground states\cite{HartmannNowak},
Swift, {\em et al},\cite{Swiftetal}
find $\beta=0.025\pm 0.015$, and
Rieger\cite{RiegerRFIM}
found $\beta \approx 0$ at positive
temperature, but without any latent heat or multipeak structure
in the magnetization distribution, suggesting
a second order transition.
However, we can more clearly exclude $\beta/\nu = 0$ as a possibility
by making use of connections
between the value of $\beta/\nu$ and the
statistics of spin clusters.

One question that naturally arises concerns the structure of spin
clusters for small $|m|$ near the critical point, where the sample
no longer {\em appears} ferromagnetic.
Is there a possibility of percolation of both up and down spins, when
$|h-h_c| \stackrel{\textstyle<}{\sim} 10^{-(22\pm 8)}$ in large enough
samples?
For fixed $+$ or $-$ boundaries,
as $h\rightarrow h_c$, $|m|$ becomes small, but the minority and majority
spins are not independent. Hence, even though the density of $+$ and $-$ spins
becomes almost equal, the minority spins are large clusters
embedded within the matrix of majority
spins, so that {\it only one sign of spin
percolates in the disordered phase} even close enough to
the critical point that the magnetization is very small and the
density of minority spins almost one half.
Exactly at the critical point in an infinite sample,
or where $\xi > L$ in a large finite sample,
the long length scale characterization of the spin
configuration will be rather different than in either phase;
these differences motivated some aspects of the present numerical study. 

\subsection{Number of states}

To study the RFIM phases and transition in more detail,
we have analyzed the influence of 
boundary conditions on a window of size $w$ in the center of a sample
as the sample size $L$ diverges.
As made clear by Newman and Stein \cite{NewmanStein},
the character of the thermodynamic limit of the
ground states can be investigated by studying such windows.
Our
numerical computations strongly support the picture of a
small number of ground states --- two in the ordered phase and one in
the disordered phase ---
consistent with the simple scaling
scenario.\cite{FisherHuseStates,BrayMoorestates}

Nevertheless, because $\beta/\nu$ is so small, at the critical point
it it is difficult to use numerics to directly distinguish between two
scenarios: (A) two coexisting states, as in the ferromagnetic phase; or
(B) a single state, with interior spins unaffected by boundary
conditions as $L\rightarrow\infty$.  If $\beta$ were exactly zero, as
in (A), then the probability $q$ that boundary conditions could affect
spins in the center in ways other than the apparent ``up" and ``down"
phases would decay as a power of the system
size,\cite{AAMstates,PalassiniYoung2d} $q\sim L^{d_s - d}$, where
$d_s$ is the fractal dimension of domain walls.  In scenario (B), a
similar power law scaling is expected, with $q\sim L^{d_s - \beta/\nu
- d}$, where the change in exponent reflects the freezing of spins or,
equivalently, decay of magnetization, as $L\rightarrow\infty$. As will
be argued below, the simplest expectation is that the exponent $d_I = d_s -
\beta/\nu$, so that $q \sim L^{d_I - d}$.
This is consistent with the assumption of only one state at criticality.

\section{Outline}

The remainder of the paper gives the details of the numerical results
and related scaling arguments. Table\ \ref{numtable} is a summary of
the numerical values of the exponents.
In \secref{sas}, we describe how the stiffness is computed and demonstrate that its scaling is quite consistent with a ``conventional" second order phase transition.
To aid in developing understanding, we study how the probability that the stiffness is exactly zero depends on
the sample shape.
\secref{dwall} presents the three methods that we employ to compute
the dimension of the domain walls generated by comparing different boundary
conditions (the same comparison used when calculating the stiffness.) The
methods differ somewhat in how they count regions of
``frozen'' spins that are not affected by boundary conditions.
In the subsequent section (\secref{mag}), we report results on the
magnetization $m$ near $h_c$. Though the distribution of $m$ depends strongly
on boundary conditions, the scaling of these
distributions are quite consistent with a single value of $h_c$ (and also
consistent with the methods of finding $h_c$ in other sections.)
Our study of the scaling of the surfaces of spin clusters with their volumes
is summarized in \secref{clust}. Besides giving a fractal dimension
$d_s^c$ consistent with the domain wall dimension $d_s$, these computations
can be used to separately infer $\beta/\nu$, given an understanding of
magnetization and correlation functions based upon a domain wall picture.
Our estimates for the singular behavior of the specific heat are included
in this section.
The general scaling picture that connects these results is reviewed in
more detail in \secref{scale}.
In \secref{gs}, we report results of how the spin configurations
depend on sample size and boundary
conditions for a fixed disorder realization.
These results are consistent with a single transition separating
a (large $h$, disordered) phase with a single
thermodynamic limit from a (small $h$, ordered) phase with two distinct thermodynamic
limits.
In the Summary (\secref{summ}),
we review the scenario for the transition that is consistent with the
numerical results and contrast this scenario with alternate
pictures.

\begin{table*}
\caption{Table of numerical estimates. The exponent or constant name,
computed value, primary method for inferring the value, section discussed,
and most relevant figure are listed.}
\label{numtable}
\begin{tabular}{|ccl|}\hline
Symbol & Value & Definition and data used \\
\hline
$\theta$ & $1.49 \pm 0.03$ & Scaling of stiffness at $h_c$, violation of
hyperscaling.\\
& & Found from scaling of stiffness with $L$ and $h-h_c$,
see \secref{sas} and \figref{fig_tension}.\\\hline
$h_c$ & $2.270 \pm 0.004$ & Critical value of the random field.\\
& & Determined from constancy in $P_0$, probability of zero stiffness
(see \figref{fig_Psigma}
and \secref{sas}),\\& & and consistent with estimates from convergence of
effective dimension estimates $\tilde{d}_{s,I,J}$,\\&&
scaling of $\Delta_{m^2}$ peak locations with $L$, scaling of $|m|$ with $L$
(\figref{fig_meanabsmag}),\\&&and window change
probabilities (\figref{fig_ofrf}).\\\hline
$\nu$ & $1.37 \pm 0.09$ & Correlation length exponent.\\
& & Found from scaling of the stiffness with $L$ \secref{sas} and
\figref{fig_tension}, with $h_c$ fixed by $P_0$ measurements.\\
&&Consistent with scaling of $\Delta_{m^2}$ peak locations with $L$.\\\hline
$\zeta$ & $0.66 \pm 0.03$ & Roughness of domain walls in the
ferromagnetic phase.\\
& & Found using anisotropic scaling (\figref{fig_scalerough}) and
effective exponent
in $L^{2/3}\times L^2$ samples.\\
& &See \secref{secrough}.\\\hline
$d_s$ & $2.30 \pm 0.04$ &
Fractal dimension of connected domain wall at $h=h_c$.\\
& & Found from surface of ${\cal U}_{++,+-}$, as shown in
\figref{fig_walls_define}.\\
& & See \figref{ds_connected} and \secref{secds}.\\\hline
$d_I$ & $2.24 \pm 0.03$ & ``Incongruent'' fractal dimension of domain wall at criticality.\\
& & Box counting of incongruent volumes (disconnected wall). See \figref{fig_d_box} and \secref{secboxcount}.\\
& & Consistent with scaling of state overlap probabilities shown in \figref{fig_ofrf}.\\\hline
$d_J$ & $2.18 \pm 0.03$ & Energy ``fractal dimension'' at $h=h_c$.\\
& & Found from the exchange part, $\Sigma_J$, of the stiffness.\\
& & See \figref{lnder_Twistbond} and \secref{secdJ}.\\\hline
$d_s^c$ & $2.27 \pm 0.02$ & Fractal dimension of the surface of spin clusters.\\
& & See \figref{fig_clustersurface} and \secref{secsurf}.\\\hline
$\rho_{\infty}$ & $0.0019 \pm 0.0004$ & Probability per scale $e$ of crossing a
spin cluster surface at $h=h_c$.\\
& & See \secref{secrho} and
Figs. \ref{fig_clustercount}, \ref{fig_clustercount},
and \ref{fig_clustercount_cf_h2}.\\\hline
$\beta/\nu$ & $0.011 \pm 0.003$ & Ratio of magnetization exponent to $\nu$.\\
& & Determined from $\rho_{\infty}$ and consistent with scaling of
$|m|$ vs. $L$ at criticality.\\
& & See \figref{fig_meanabsmag} and \secref{seccritcorr}.\\\hline
$\beta$ & $0.017 \pm 0.005$ & Magnetization exponent, found from $\beta/\nu$ and $\nu$.\\\hline
$(\alpha -1)/\nu$ & $-0.74 \pm 0.02$ & Combination of heat capacity
exponent $\alpha$ and $\nu$\\
& & Found using value for $d_s^c$ and \eqref{clustEJ}.\\\hline
$\alpha$ & $-0.01 \pm 0.09$ & Heat capacity exponent,
found using \eqref{clustEJ} and $\nu$. \\
& & Consistent with modified hyperscaling \eqref{hyper}
and the value $\alpha=-0.12
\pm 0.12$ found\\
& & from a fit to the bond energy density $E_J(L)$ at $h_c$ and $\nu$.\\\hline
\end{tabular}
\end{table*}

\section{Stiffness and  scaling}\label{sas}

To establish the location and nature of the transition, we first focus
on the stiffness of the system. In an ordered Ising phase, the (free)
energy of a domain wall across a system of size $L^d$ will be
$\Sigma\approx \sigma L^{d-1}$ with $\sigma$ the interfacial
tension.  At an ordinary first order
transition, the interfacial tension is discontinuous at
the transition, while near a second order transition, it
goes smoothly to zero. For a zero-temperature transition, the interfacial tension vanishes 
with a variant of Widom scaling \cite{FisherRFIM}
\be
\sigma \sim (h_c-h)^{(d-1-\theta)\nu} .
\ee

To probe the stiffness of a random system takes some care. For a
random field system, there is no exact symmetry between the up and the
down spins but only a {\it statistical symmetry of the distribution}
of the random fields.  Thus, for example, for a given sample in the
ordered phase, the energy of the up state --- obtained with up
boundary conditions (BCs) --- will differ from that of the down state (obtained from down BCs), by a random amount of order $\sqrt{\cal{V}}$
arising from the differing effects of the random fields on the two
states.  In order to compute the interfacial energy it is useful to
subtract as much as possible of this random ``bulk" energy so as to be
left with a quantity that is as-close-as-possible to an
``interfacial" energy.

\subsection{Definition of the stiffness}

To obtain the stiffness of a sample, we compute the {\it symmetrized
energy difference} between antiparallel and parallel boundary
conditions. This is computed from the ground state energies for four
different boundary conditions on a given sample, denoted $++$, $+-$,
$-+$ and $--$. These correspond to fixing the spins to have values
$s=+1$ or $s=-1$ on the left or right sides while imposing periodic
boundary conditions in the other two directions.  For example, $+-$
has spins fixed to $+1$ on the left and to $-1$ on the right.  The
interface energy is then defined as \cite{FisherHuseStates}
\be
\Sigma \equiv (E_{+-}+E_{-+}-E_{++}-E_{--})/2.
\ee
Note that the average over samples of $\Sigma$ will be the same as
that of $E_{+W} \equiv E_{+-} - E_{++}$.  Studying $\Sigma$, however,
reduces the effects of energy changes near the boundaries that are
caused by the differing boundary conditions: in $\Sigma$, each
boundary condition on each side appears twice but with opposite signs
so that these effects cancel.  This cancellation will be most
pronounced well into the disordered phase.

In the {\it disordered phase}, the boundary conditions typically only affect
layers near the boundaries with thickness of order the correlation
length $\xi$; deep in the interior (for system sizes $L\gg \xi$) the
spins will be frozen, completely unaffected by the boundary
conditions.  The {\it average} energy of the boundary layers will,
because of the statistical symmetry, be independent of whether the
boundary conditions are plus or minus. But there will be a random part
of the boundary energy, with magnitude of order $\sqrt{L^{d-1}}$, that
{\it is} sensitive to the boundary condition.  Thus in three
dimensions, $E_{+W}$ will typically be of order $L$ even in the
disordered phase.  In contrast, the stiffness $\Sigma$ will typically
be exactly zero because of the cancellation of the boundary energies
and the concomitant frozen interior which blocks any knowledge
of the spins near
one face about the boundary conditions on the opposite face.
In general, 
the distribution for $\Sigma$ contains a $\delta$-function
contribution with some weight \be P_0 \equiv \pr[\Sigma=0].
\ee

\begin{figure}
\centering
\includegraphics[width=6.5cm]
{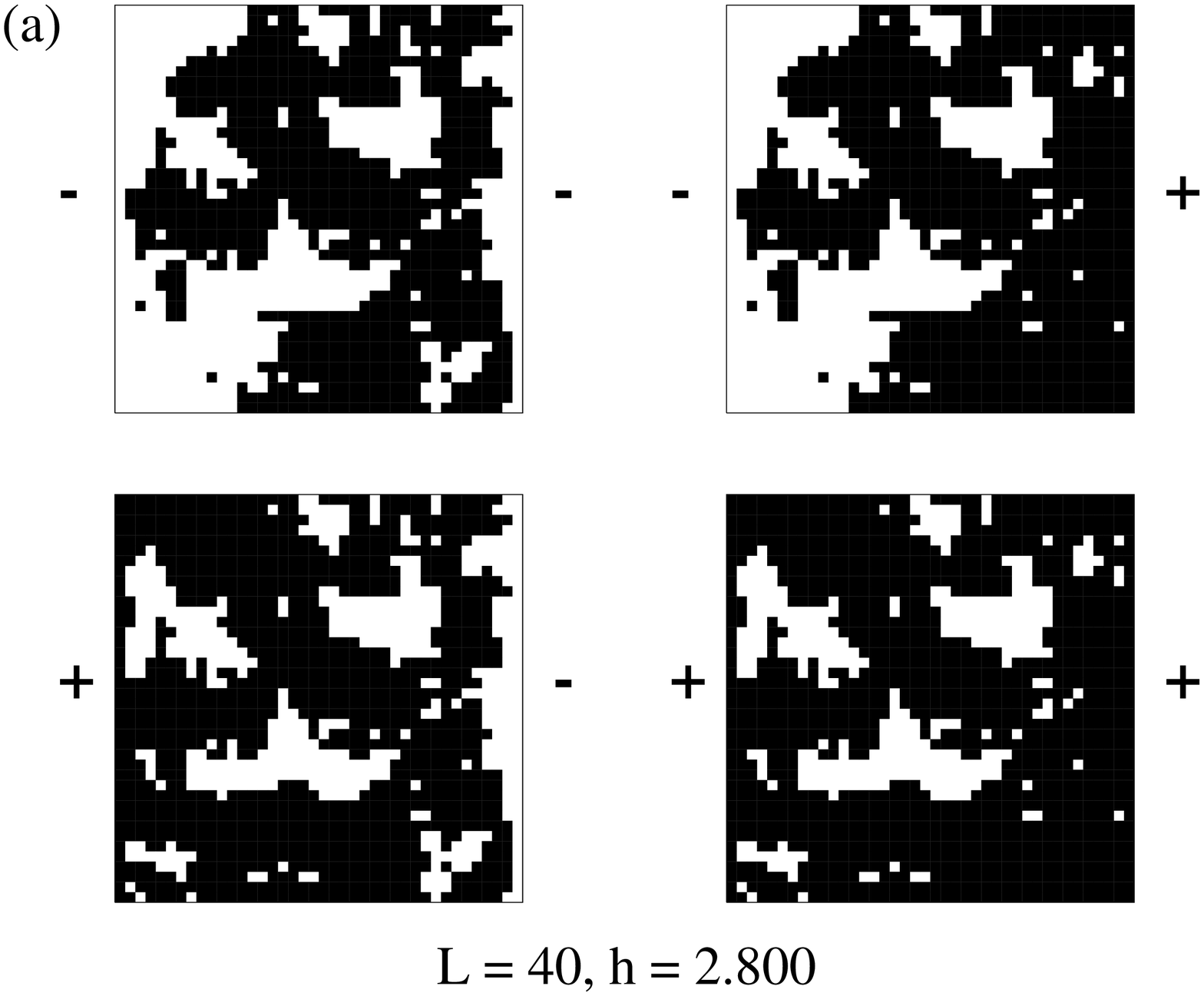}\\
\centering
\includegraphics[width=6.5cm]
{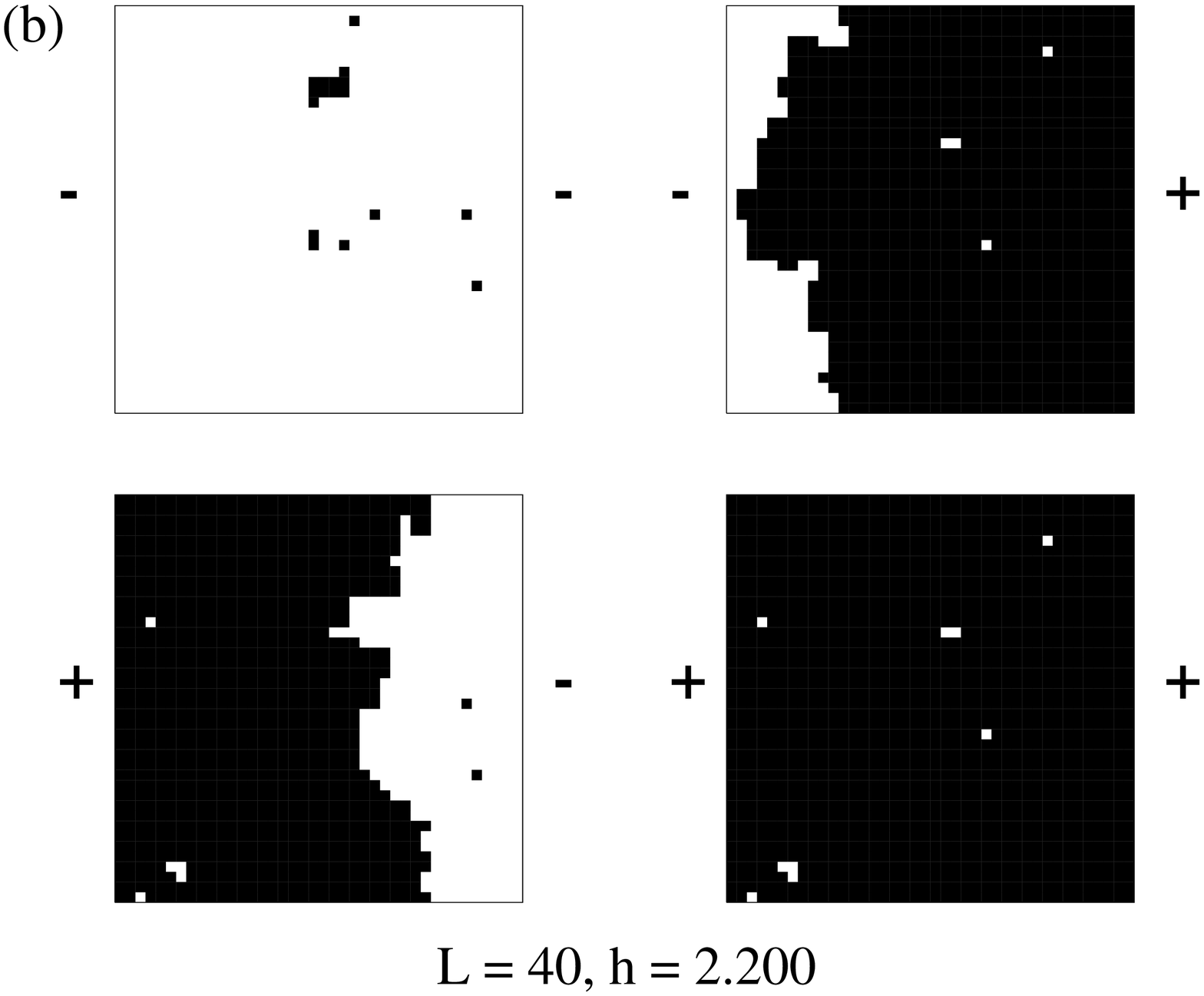}\\
\centering
\includegraphics[width=3.5cm]
{1c.eps}
\includegraphics[width=3.5cm]
{1d.eps}
\caption{
Pictures of planar slices  ($z=0$ ) of configurations,
for fields (a) $h=2.8$ and (b) $h=2.2$, in samples of size $40^3$.
The slices shown at each $h$ are  the four combinations of boundary
conditions
$--$ (top left),
$-+$ (top right), $+-$ (bottom left), and $++$ (bottom right), where the
left and right faces (in the $x$ direction) are fixed $+$ or $-$ and
periodic boundary conditions are in effect for the $y$ (up/down) and
$z$ (out of the page) directions.
The dark squares indicate an up spin at that location in the slice.
The $--$ and $++$ visualizations
for $h=2.2$ show the presence of minority spin ``bubbles'' embedded
in the bulk.
A summary of the effect of the boundary conditions for $h=2.8$ and
$h=2.2$ are shown in parts (c) and (d), respectively.
Dark and light
squares correspond to up and down spins,
respectively, that are {\it frozen},
i.e., invariant under this set of boundary conditions.
The gray {\em controllable}
spins can be modified by choosing among the four boundary
conditions.
For $h=2.8$, the gray volume is composed of two unconnected regions
anchored on the two controlled boundaries,
so that the stiffness $\Sigma=0$.
In contrast, at $h=2.2$, in the sample shown, the gray
region connects the two sides and $\Sigma \ne 0$.
}
\label{fig_twistpics}
\end{figure}

Sample configurations from simulations
are illustrated in \figref{fig_twistpics} with
two-dimensional slices shown.
Part (a) of the figure illustrates
a situation somewhat into the disordered phase in which the left and
right boundaries are effectively decoupled as discussed above.  The
{\it frozen spins}, those that are the same with all four boundary
conditions, are indicated by dark or white squares
in part (c) of the figure,
while those that are affected by the BCs, the controllable
spins, are
indicated by gray squares.

The behavior in the {\em ordered phase} is quite different as can be seen in
the parts (b) and (d) of \figref{fig_twistpics}. In this case, the difference
between the $--$, the $+-$ and the $++$ boundary conditions can be
well characterized by a $+|-$ domain wall that has a minimum energy
position somewhat to the left of the center. Similarly the difference
between the $--$, the $-+$ and the $++$ boundary conditions is
characterized by a $-|+$ domain wall whose minimum energy position is
somewhat to the right of the center.  The stiffness of this sample
will thus be half the sum of energies of the two types of walls {\it
plus} the energy of the random fields (here predominantly negative) in
the region between the two favored positions of the walls; the
contribution of the random fields in this region will {\it not}
cancel.  

This picture yields a stiffness in the {\it ordered phase} with a
mean of order $L^{d-1}=L^2$ and variations around this mean of order
$L^{d/2}=L^{\frac{3}{2}}$, the variations being dominated by the
random fields in between the positions of the two types of walls.
In the ordered phase for $h<h_c$, $P_0\rightarrow 0$ --- apparently
exponentially fast or faster in $L$ ---
as $L\rightarrow\infty$.
 
At the {\it critical point}, the behavior is qualitatively like that in the
ordered phase.  But here the energy cost of the interface is much
lower, the interface itself is fractal, and,
in the regions of controllable spins
that are otherwise flipped by
the changing boundary conditions,
there are large
frozen unflipped ``holes".

\subsection{Statistics of the stiffness}

The stiffness $\Sigma$ was determined by finding the ground
states for a single sample subject to each of the four boundary
conditions $++$, $--$, $-+$ and $+-$.  By studying many samples, we computed the distribution of $\Sigma$ for various system sizes
and random field strengths. In particular, we computed both the
probability $P_0$ that $\Sigma=0$ as well as
the mean stiffness $\overline{\Sigma}$, denoting, as usual, averages over the
randomness by overlines. For the bulk of the computations, ground
states were found for cubic samples of size $L^3$ and anisotropic
samples of size $2L \times L^2$ with the length along a
$(100)$ axis of the lattice perpendicular 
to the controlled faces, the $x$-direction, being $2L$.
To check that the results were not
artificially influenced by lattice orientation
effects, we also computed
values of $\Sigma$ for two types of samples whose
controlled faces are $L\times L$
rhombi, with $L$ and $4L$ layers, respectively,
separating the two faces along the
$(111)$ direction.  Note that such $(111)$ layers are separated by a
distance of $1/\sqrt{3}$, rather than the distance of $1$ that separates
the $(100)$ layers. As the lattice in both of these geometries is
cubic and the ferromagnetic couplings are the same, the $h_c$ found
should be the same in the two orientations.

\begin{figure}
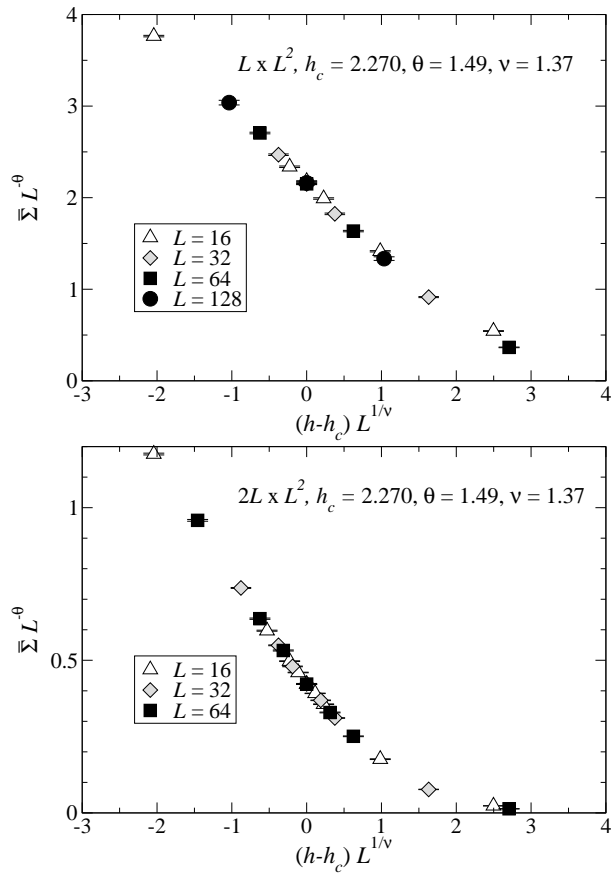

\centering
\includegraphics[width=8.0cm]
{2a.eps}
\centering
\includegraphics[width=8.0cm]
{2b.eps}
\caption{
Scaling plot for the sample averaged stiffness $\overline{\Sigma}$
for (a) isotropic samples of volume $L^3$ and
(b) anisotropic samples of volume $2L \times L^2$, with the longest
axis being the direction in which the boundary conditions are varied.
Note that the vertical scales differ.
The stiffness is calculated by the symmetric comparison of
four ground state energies: the energies for the four choices of
spin up and spin down boundary
conditions on the left and right sides and with periodic boundary conditions
in the other two directions.
The fit shown is for energy exponent $\theta = 1.49$, correlation length
exponent $\nu = 1.37$ and critical value $h_c = 2.270$. This scaling is
consistent within errors, except for the $L=16$ isotropic
samples.
Statistical ($1\sigma$) error bars are shown.
}
\label{fig_tension}
\end{figure}

If the transition is second order, the mean
interface energy should scale as
\be\label{sigma-bar-scaling}
\overline{\Sigma} \approx \overline{C}L^{\theta} {\cal S}[L^{1/\nu}(h-h_c)K],
\ee
where the exponent $\theta$ sets the scaling of the energy at the
critical point, $\nu$ is the correlation length exponent, ${\cal S}$ a
universal scaling function that depends on the {\it shape} of the
sample, and $K$ and $\overline{C}$ are non-universal coefficients.  Using this scaling form
and varying $\theta$, $\nu$ and $h_c$ yields a good collapse of the
data, as shown in \figref{fig_tension} for the $(100)$ samples.
By varying the exponents in
the scaling plot, we estimate the values $h_c = 2.27 \pm 0.01$,
$\theta = 1.50 \pm 0.08$ and $\nu = 1.35\pm 0.20$.
Using the data for $P_0$ to fix $h_c =2.270$
as discussed below, gives $\theta = 1.49 \pm 0.02$
and $\nu = 1.37 \pm 0.09$.
The excellent collapse of the data strongly support the conclusion that
the phase transition is second order.

The value of $\theta$ is in quantitative agreement with results from
Monte Carlo simulations at finite temperature,\cite{RiegerRFIM} which 
found $\theta=1.53\pm 0.10$.
It is also within the bounds determined from various arguments:
\be
\frac{d}{2} -\frac{\beta}{\nu} \le \theta \le \frac{d}{2};
\ee
the lower bound arising from scaling laws and a rigorous inequality
\cite{FisherRFIM,SofferSchwartz}. The upper bound follows from the observation
that any larger value of $\theta$ would imply that the system would be
stable --- by the argument of Imry and Ma \cite{ImryMa} ---
to an increase of the random
field and thus should not be at the critical point. Since $\beta/\nu$
is extremely small, we expect that the true value of $\theta$ should
be just slightly below $3/2$.  This is to be contrasted with the
``dimensional reduction'' result predicted to obtain to all orders in
a $d=6-\epsilon$ expansion of $\theta=2$ (but see recent claims in
Ref.\ \onlinecite{BrezinDeDominicis}).

The correlation exponent $\nu$ must be no smaller than $2/d$ in random
systems.\cite{ChayesChayesFisherSpencer}
Our result easily satisfies this bound.  Indeed, it is
substantially larger than this lower bound and even more so than the
mean-field value of one half; this is presumably associated with
proximity to the lower critical dimension of $d_\ell =2$.

\subsubsection{Stiffness in the disordered phase}
\begin{figure}
\centering
\includegraphics[width=8.0cm]
{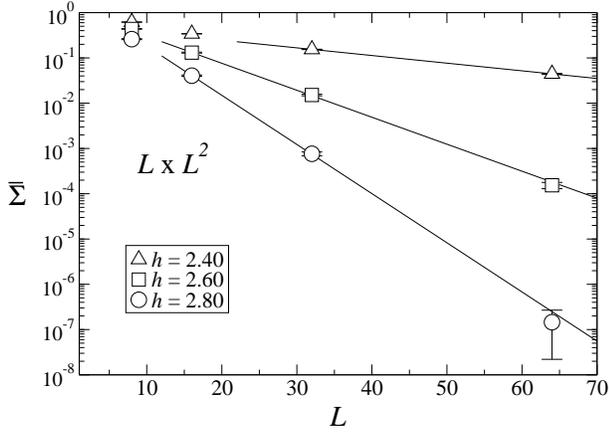}
\caption{
Plot of the decay of the mean stiffness
$\overline{\Sigma}$ with $L$, in the disordered phase.
The lines are fits to the data for $\overline{\Sigma} < 0.2$ of the form
$\overline{\Sigma}\sim \exp(-L/\xi_\Sigma)$. In conjunction
with similar fits for $2L\times L$ samples, which allow us to
estimate errors from finite size fitting, 
we find the values $\xi_\Sigma = 26 \pm 4$, $7 \pm 1$
and $4.0 \pm 0.4$ for $h=2.4$, $2.6$, and $2.8$, respectively.
}
\label{fig_stiffvsL}
\end{figure}

\figref{fig_stiffvsL} shows the dependence of the mean stiffness on
the linear dimension $L$ for the $L \times L^2$ samples.
The decay of the stiffness
is well fit by a decaying exponential
$\overline{\Sigma} \sim \exp(-L/\xi_\Sigma)$, for $h>h_c$ and
$L >> \xi_\Sigma$ (roughly when ${\overline{\Sigma}} < 0.2$.)
The correlation length can be inferred from the fits. The values for
$\xi_\Sigma$
obtained from the $2L \times L^2$ samples, using a similar
plot, are in agreement with those from the cubic
sample to within $10\%$ for each $h$.
The values of the correlation lengths $\xi_\Sigma$
found are consistent with a divergence of
$\xi_\Sigma \sim (h-h_c)^{-1.3\pm0.1}$, taking $h_c=2.27$,
consistent with our other determinations of $\nu$, though the
data is not very near $h_c$.
It may be possible to make a more accurate
determination of $\nu$ by more careful calculations using $h$
values somewhat nearer to $h_c$.

For the $2L \times L^2$ samples, $P_0$, the probability
of the stiffness being zero, can be appreciable for accessible
$L$ and $h$ near $h_c$.
The data for fixed $h \ge h_c$ are consistent with $P_0$ approaching
one exponentially with $L$, although other forms cannot be
ruled out.

\subsubsection{Stiffness at criticality}

\begin{figure}
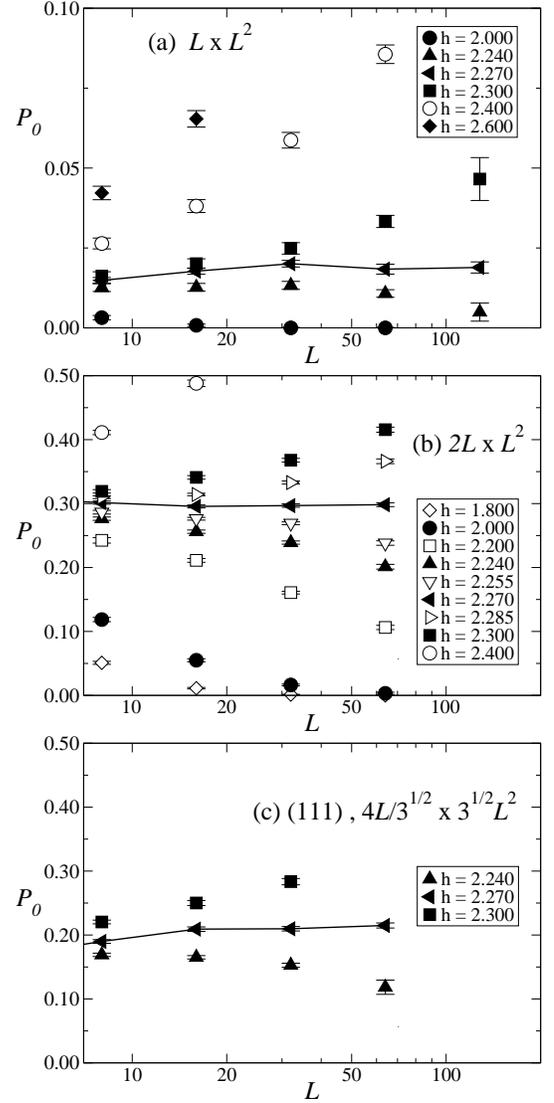

\centering\includegraphics[width=7cm]
{4a.eps}
\centering\includegraphics[width=7cm]
{4b.eps}
\centering\includegraphics[width=7cm]
{4c.eps}
\caption{
Plot of the probability $P_{0}(h, L)$ that the stiffness
$\Sigma$ is zero,
for (a) isotropic samples of volume $L^3$ and
(b) anisotropic samples of volume $2L \times L^2$, with the longest
axis being the direction in which the boundary conditions are varied,
and (c) anisotropic samples of volume $4L^3$, with the boundary faces
in the (111) plane.
For all sample shapes, the convergence to a fixed value of $P_0$
as $L\rightarrow\infty$ for $h=2.270$ suggests the location of the
critical point.
The solid lines connect the points for $h=2.270$ to demonstrate convergence
of $P_0$ to a constant, within statistical errors.
As $P_{0}$ is very nearly zero for isotropic samples
($P_{0}(2.27,\infty)\approx 0.04$, if the apparent convergence holds
at large $L$),
the errors in determining $h_c$ are larger. From the
$2L \times L^2$ anisotropic samples,
where the apparent extrapolation is $P_{0}(2.27,\infty)=0.298\pm 0.05$,
$h_c = 2.270\pm 0.004$.
For the (111) oriented samples, with volume of
$4L/\sqrt{3} \times \sqrt{3}L^2$ (layer separation $\times$ layer area),
the data is also consistent with
$h_c = 2.270$, with $P_0=0.23 \pm 0.01$.
Separate results, not shown, for (100) samples of shape $4L \times L^2$ give
a value of $P_{0}(2.27,16 \le L \le 64)=0.79 \pm 0.02$.}
\label{fig_Psigma}
\end{figure}

At the critical point, a non-trivial scaling function
\be
P_c(\Sigma/CL^\theta)\equiv P(\Sigma/CL^\theta,0)
\ee
 for the
distribution of $\Sigma$, would suggest that $P_0$ should approach a
{\it finite fixed point value}.  In \figref{fig_Psigma}(a) a plot of
$P_0$ is shown as a function of $L$ for various $h$.  Observe that
opposite boundaries are almost always coupled at
$h_c$ in the cubic samples:
\be
P_0^{\rm cubic}(h_c) \simeq 0.04 \pm 0.01
\ee
This value for $P_0$ is so small
that to verify that $P_0$ indeed
approaches a non-zero constant at the transition, we also performed
simulations for anisotropic samples of various shapes.
In general, we expect that the distribution of
$\Sigma$ at the critical point, $P_c$, will depend on the {\it shape}
of the sample with long thin samples typically yielding
lower stiffness and a higher probability of the stiffness vanishing
than short fat ones of the same cross section. 

For rectilinear samples of dimensions $2L\times L^2$, with
the controlled boundaries at opposite ends of the {\it long}  $(100)$ axis, we find that $P_0$ approaches
a value well away from zero, $P_0 = 0.298 \pm 0.005$, at
$h=2.270$
(\figref{fig_Psigma}(b)).
For rhomboidal samples with $L^3$ spins consisting of $L$ layers
and a length along the $(111)$ control axis of $L/\sqrt{3}$, $P_0$ is not
distinguishable from zero. However, for longer rhomboidal samples with $4L^3$ spins consisting of $4L$
layers and a length $4L/\sqrt{3}$ along the control axis, we find $P_0 = 0.21 \pm 0.01$ at $h=2.270$ for
$16 \le L \le 64$ (\figref{fig_Psigma}(c)).
Imposing convergence of $P_0$ to a fixed
(non-trivial) value as $L\rightarrow\infty$ gives a critical value of
$h_c=2.270$.  These anisotropic rectilinear
and long rhomboidal samples yield our most precise estimate for $h_c$,
Eq.(\ref{hc}).  

We  should expect  
$P_0$ to be a smooth function of the shape; the fact that it is
far from zero in samples with aspect ratio of order two lends strong support to
the conjecture that it will be non-zero for any shape.  The
observation that it is small for cubical samples is related, as will
be explained below, to the smallness of $\beta$.

\subsubsection{Comparison of distributions for $\Sigma$}
\begin{figure}
\centering\includegraphics[width=8.cm]
{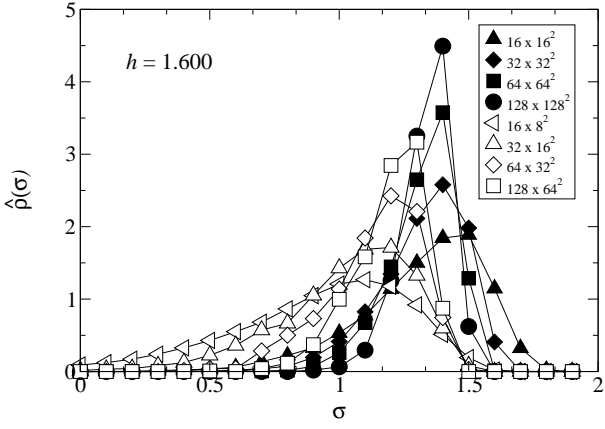}
\caption{
Probability density $\hat{\rho}(\sigma)$ of the values of the interfacial
energy density $\sigma = L^{-2}\Sigma$
in the ordered phase with $h=1.60 < h_c$.
As the sample size grows larger, the relative
sample-to-sample variations of $\sigma$
decrease, consistent with an approach to a $\delta$-function
at the mean value $\sigma(h=1.6)\approx 0.69$,
for both the $L^3$ and $2L\times L^2$ samples,
}
\label{fig_twist160}
\end{figure}

\begin{figure}
\centering\includegraphics[width=8.cm]
{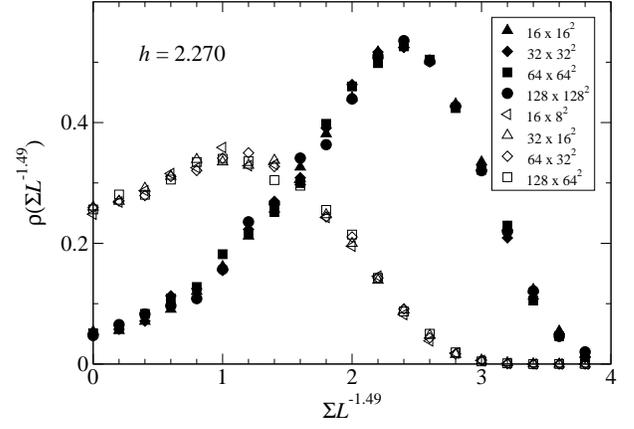}
\caption{
Probability density $\rho(\Sigma)$ of the scaled
non-zero values of $\Sigma L^{-\theta}$ with $\theta=1.49$,
for $h=2.270 \approx h_c$. A $\delta$-function
at $\Sigma = 0$ with weight $P_0\approx 0.04$ ($\approx 0.298$)
for the $L^3$ ($2L \times L^2$) samples is not shown.
The distributions for samples with linear dimensions $L$
greater than $16$
are statistically
consistent with a fixed point distribution for $\Sigma$, with a
characteristic scale $\Sigma_0 \sim L^{1.49}$ and a form
dependent on the shape of the samples.
}
\label{fig_twist227}
\end{figure}

\begin{figure}
\centering\includegraphics[width=8.cm]
{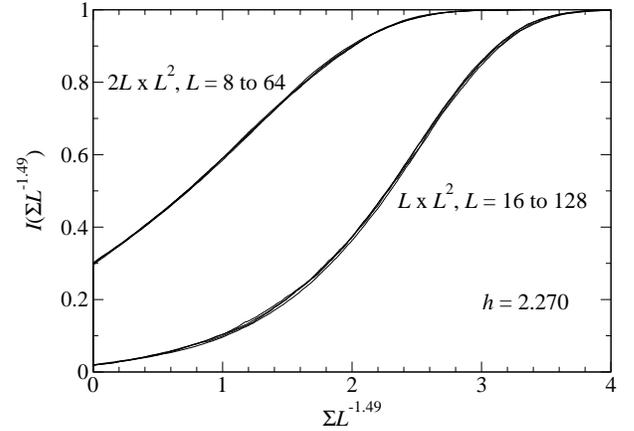}
\caption{
Scaling plot for the
cumulative distribution $I(\Sigma) = \int_0^\Sigma d\Sigma' \rho(\Sigma')$ 
of the stiffness,
for $h= 2.270 \approx h_c$.
The stiffness $\Sigma$ has been
scaled by the energy scale $L^{\theta}$, with $\theta = 1.49$.
The labels indicate the sample shapes ($2L\times L^2$ and
$L \times L^2$) for each set of curves.
For each sample shape, four sample sizes are plotted
($16\times 8^2 \rightarrow 128 \times 64^2$ and $16^3\rightarrow
128^3$.) At the resolution shown, the scaled curves are nearly
independent of $L$.
The intercept at $\Sigma = 0$ corresponds to $I(0) = P_0$,
the probability of
a sample having zero stiffness.
As in \figref{fig_twist227},  the curves converge to a fixed point
distribution.
}
\label{fig_cum227}
\end{figure}

\begin{figure}
\centering\includegraphics[width=8.cm]
{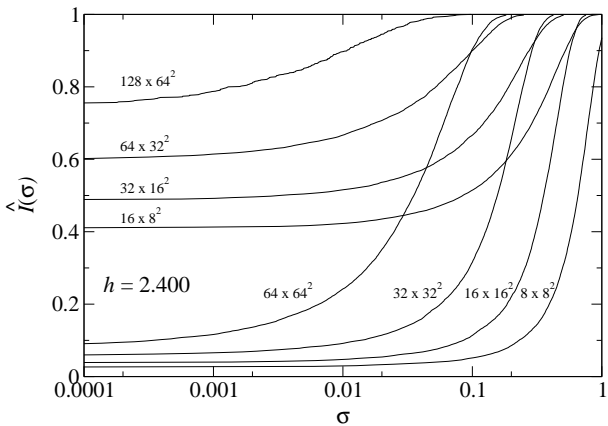}
\caption{
Cumulative distribution $\hat{I}(\sigma)
= \int_0^\sigma d\sigma' \hat{\rho}(\sigma')$ 
for $h=2.400 > h_c$.
As the sample size increases, $P_0$, given by the intercept of
the curves at $\Sigma=0$, increases, and the typical
non-zero
values  of $\sigma\equiv\Sigma L^{-2}$ rapidly decrease.
From \figref{fig_stiffvsL},  the length scale for the
decay of the stiffness is $\xi_\Sigma = 26 \pm 4$. This length is
comparable to the mid-range
system sizes here. Note that $P_0$ rises more quickly
and the typical non-zero $\sigma$ decays more rapidly with $L$ in the
anisotropic samples.
}
\label{fig_twist240}
\end{figure}

The complete probability distributions for $\Sigma$ at various values
of $h$ are plotted in Figs.\ \ref{fig_twist160} through \figref{fig_twist240}
for both the cubical and the elongated $(100)$ samples. For $h=1.6$, the
distribution of $\Sigma$ appears to approach a narrow distribution
about $\Sigma \approx (1.39) L^2$, regardless of the sample shape; this
is as expected for the ordered phase.  For $h=2.27$, the critical
point, the distribution obeys the simple scaling form of
Eq.(\ref{sigma-scaling}) for both the isotropic and anisotropic
samples but with a different scaling functions for each of the two
shapes.  \figref{fig_twist240} shows the integrated probability
distributions for $h=2.40$. As $L$ increases, the mean interfacial energy
decreases approximately exponentially, and $P_0$ approaches $1$.

\section{Geometry of Domain walls}\label{dwall}

In addition to the scaling properties of the energies of domain walls,
we are also interested in their geometrical properties. These properties
are expected to be related to the properties of the surfaces of spin
clusters that are either frozen or induced by bulk perturbations (as opposed
to boundary perturbations), to the effects of boundary conditions on the deep
interior of a sample, and to the general scaling picture for the transition.

In the
ferromagnetic phase, the interfacial tension $\sigma$ is positive.
The domain walls will appear flatter and flatter
on large length scales with surface area proportional to $L^2$ but
nevertheless divergent roughness characterized by a roughness
exponent $\zeta = 2/3$ and random energy variations
that scale as $L^{\theta_I}$, with
$\theta_I = 4/3$.\cite{GrinsteinMa,HuseHenley,FisherFRG,ChauveLeDoussalWiese}

At the critical point, the interfacial tension of the walls
vanishes. Thus we should not except them to be flat even on large
scales; the natural expectation is that they will be fractal with
surface area scaling as
\be
A\sim L^{d_s} 
\ee
with $d_s$ a {\it fractal dimension}. One might expect, {\em a priori},
that the exponent $d_s$ would be an independent exponent as it is not
obviously related to $\theta$, $\beta$, and $\nu$.  For the simple
scaling scenario to obtain, we expect $d_s$ to be in the range:
\be
d-1<d_s<d .
\ee
If the transition were first order, one would expect $d_s=d-1$ as in
the ferromagnetic phase (more precisely, some fraction of samples
at the transition would show such interfaces).  If, at the other
extreme, it were found that $d_s=d$, this would mean that the ``walls"
would be spacefilling (up to possible logarithmic factors); this would
cast doubt on the overall scaling scenario for excitations, etc, near
the phase transition.\cite{FisherRFIM}

\subsection{Frozen spin regions}

To study interfaces we would like to compare the spin configurations
found using the boundary conditions $++$, $+-$, $-+$, and $--$ as
discussed in the previous section.  But in random field systems, there
is an intrinsic difficulty associated with defining an interface: this
arises from the presence of frozen regions which are not affected by
changing from $++$ boundary conditions to $--$ boundary conditions and
thus are unaffected by {\it any}
changes in the boundary conditions on the
controlled surfaces.\cite{tobepubAAM}
With mixed boundary conditions,
say $+-$, the interface between the region that is like the
``up'' $(++)$ state and  the region that is like the ``down'' $(--)$
state can
pass along the boundary of the frozen regions.  Are we to count such
sections as truly part of the interface? Or should we exclude the
frozen regions from the system and think of the interface as bisecting
only the remaining controllable regions?

\begin{figure}
\centering\includegraphics[width=8.cm]
{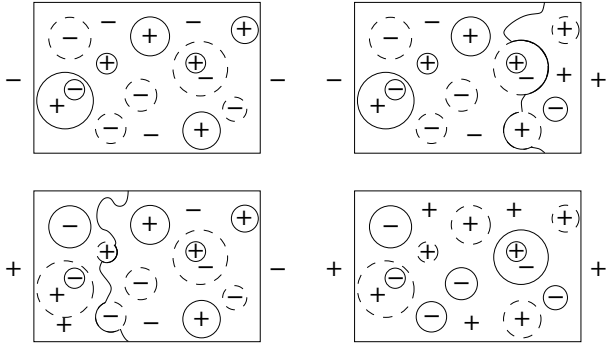}
\caption{
Schematics of the spin configurations for four different
boundary condition combinations, for a case with $\Sigma \ne 0$.  Here,
there is a set of {\it controllable spins}, connected across the sample,
that can be either $+$ or $-$, depending on the boundary conditions.
These are the majority of the spins in the figure shown. The {\it frozen
spins} are those that are constant under the four boundary conditions
$--$, $-+$, $+-$, and $++$; these are indicated here by the circular
regions.  Solid lines separate spins of opposite sign, while the
dashed lines indicate frozen islands that are of the same sign as the
surrounding spins.
}
\label{fig_walls_pic_1}
\end{figure}

\begin{figure}
\centering\includegraphics[width=8.cm]
{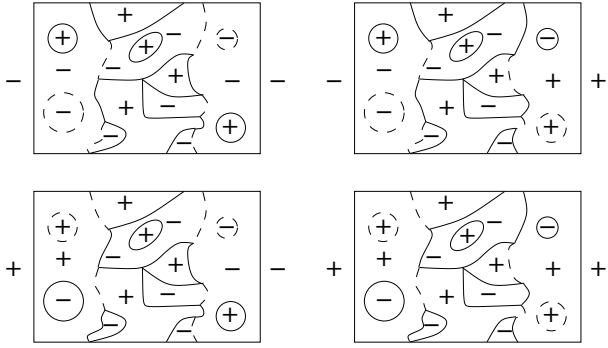}
\caption{
Schematics of the spin configurations for four different boundary
condition combinations, for a case with $\Sigma = 0$. In addition to  the
frozen islands, shown as circles as in \figref{fig_walls_pic_1},
there is a set of frozen interior spins that spans the sample in
the directions perpendicular to the horizontal (control) axis.
Conventions for solid and dashed lines are as in \figref{fig_walls_pic_1}.
The surfaces used to measure $d_s$ are the two surfaces of the frozen
interior, but
the measure used to compute $d_I$ is {\em zero}, as long as
the boxes have side $B$ smaller than the size of the frozen interior.
Also zero is the exchange
stiffness $\Sigma_J$, as each bond that is broken in both the $+-$ and $-+$
configurations is also broken in both the $--$ and $++$ configurations, while
bonds that are broken exactly once under one of the two antiparallel
BCs is likewise broken exactly once under parallel BCs,
so that all broken bonds
{\em cancel} in the signed sum that defines $\Sigma_J$.
}
\label{fig_walls_pic_2}
\end{figure}

We are thus led to consider several methods for
measuring the surface area of ``interfaces" anticipating that we might
obtain results which depend on the definition.
For these considerations, it is useful to
refer to \figref{fig_walls_pic_1}, which is a sketch of what might
happen when $\Sigma = 0$, and \figref{fig_walls_pic_2}, which is a
representation of what might happen for $\Sigma \ne 0$.

Configurations are shown for each of the four boundary combinations:
the circles enclose regions of ``frozen spins''  --- those that are
constant under all four BCs --- with solid lines
indicating broken (unsatisfied) bonds.
The dashed lines indicate the location of a frozen
cluster embedded in a set of like spins.
The interior of the configurations in
\figref{fig_walls_pic_2} are also frozen.  Note that the frozen spin
regions can contain nested subclusters of alternating spins.
In \figref{fig_walls_pic_1}, spins outside of the
frozen spin regions can be either $+$ or $-$, depending on the BC
combination.  These two figures are caricatures of configurations
such as those shown in \figref{fig_twistpics}.
Note that \figref{fig_walls_pic_1} does
not show all of the possibilities. Also,
these pictures are two-dimensional slices, which hides the
possibilities of   regions having  three dimensional
``handles"   and minimizes the potential role
of simultaneous percolation of $+$ and $-$ spins in some regions.

\subsection{Surface exponent $d_s$}\label{secds}

\begin{figure}
\centering\includegraphics[width=8.cm]
{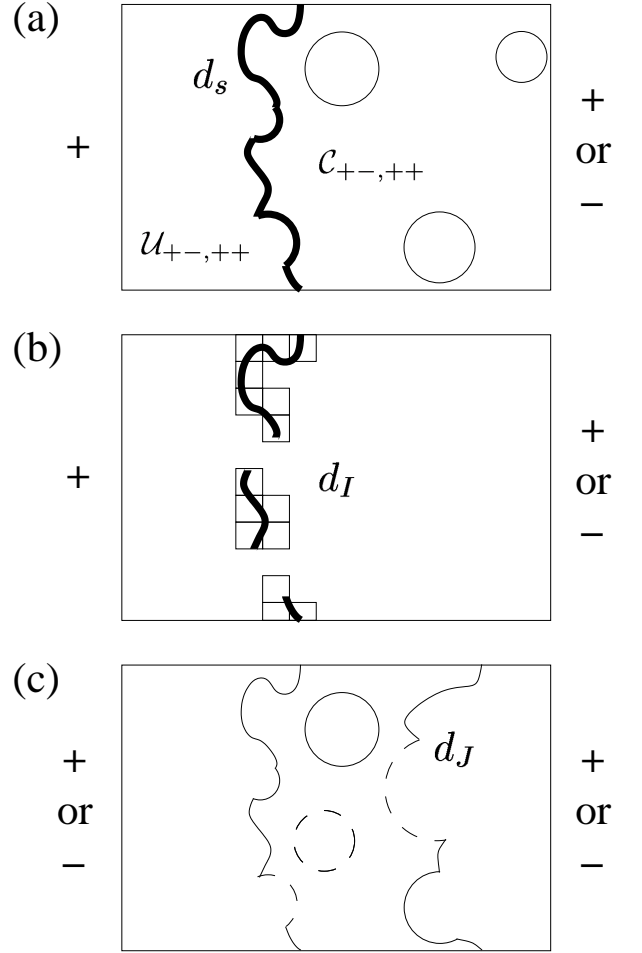}
\caption{
Schematics of the definitions of domain wall measures, based on the
configurations of \figref{fig_walls_pic_1}.
(a) The heavy solid lines indicate
the boundaries used to define the domain walls 
for the calculation of the fractal dimension $d_s$
of the spanning wall obtained
by comparing the $+-$ and $++$ configurations.
The region of
changed spins connected
to the right face is ${\cal C}_{+-,++}$, which has both the heavy and light
lines as boundary, while the unchanged connected
region anchored on the left face, with the single solid line as boundary,
is ${\cal U}_{+-,++}$.
(b) Boxes used for determining $d_I$, the dimension of the
locally incongruent regions.
The number of  boxes of side $B$ in which the $+-$
configuration differs from {\it both}
the $++$ and $--$ configurations scales as $L^{d_I}$.
The broken
bonds around the frozen islands in the $++$ or
$--$ configurations are not counted. 
(c) The signed sum of broken bonds that defines $\Sigma_J$,
the exchange contribution to the stiffness $\Sigma$. Solid lines indicate
positive contributions and the long dashed lines indicate  negative contributions.
}
\label{fig_walls_define}
\end{figure}

The first method of defining an interface uses just two different
boundary conditions, for example, the $+-$ to $++$ comparison. This
change in boundary conditions causes a connected set of spins anchored
to the right face to flip from up to down along with the forced right
boundary spins when the $++$ boundaries are replaced with $+-$. This
set of {\em changing} sites, which we denote ${\cal C}_{+-,++}$, has
a bounding surface --- indicated by the heavy and light lines
in \figref{fig_walls_define}, for the spin configurations of
\figref{fig_walls_pic_1}.
But some of this boundary will surround
islands of fixed up spins (some of which themselves have down spin
inclusions) that are disconnected from both controlled faces.
The number of such islands (light circles in
\figref{fig_walls_define})
will scale with the volume of the ${\cal C}_{+-,++}$
region and their boundaries will contribute an amount of order
this volume to the surface area of ${\cal C}_{+-,++}$.
This  internal contribution to the
surface area will, on average, be dominant in large systems when
$h<h_c$,
but it is clearly
{\it not} properly part of the domain wall.

What we are interested in is the part of the boundary of
${\cal C}_{+-,++}$ which interfaces with the other ``half" of the
system.  One way to define a domain wall is thus to start at
the unmodified {\it left} face and find the set of spins connected to
this face that do not change when the boundary conditions on the {\it
right} face are changed; this set, which we denote ${\cal U}_{+-,++}$,
is simply connected, i.e. it has no interior holes, although it could
have handles.  The surface of the set ${\cal U}_{+-,++}$ is just its
interface with the set ${\cal C}_{+-,++}$ that flips.
This ${\cal U-C}$ interface, which spans the whole cross-section
with no holes and thus includes
some boundary of frozen regions, is our first definition of a domain wall of
interest.

\begin{figure}
\centering\includegraphics[width=8.cm]
{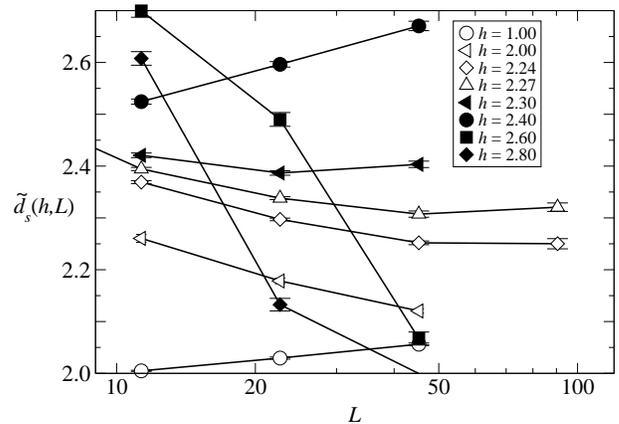}
\caption{
Effective dimensions $\tilde{d}_s(h,L)$
obtained from a logarithmic derivative of the surface area with respect
to $L$.
These are used to estimate the fractal dimension
$d_s\equiv \tilde{d}_s(h_c,\infty)$.
The values $\tilde{d}_s(h,L)$ are calculated from the surface of the
connected set of spins rooted at one face that is unchanged when
the spins on the {\it opposite} face of the sample are flipped.
The scaling of the area of this surface with $L$ yields the estimates shown,
via \eqref{eqn_dslocal}.
The error bars represent $1\sigma$ statistical uncertainties.
The values converge to $d_s = 2.30\pm 0.04$ for
$h$ near $2.27\approx h_c$,
with the error reflecting the uncertainty in $h_c$ and
the estimated magnitude of finite size corrections.
The lines connect data points with the same $h$.
}
\label{ds_connected}
\end{figure}

Averaging over samples at fixed $h$ gives a mean surface area of this
${\cal U-C}$  domain wall, $A(h, L)$.
(For these and related studies, we used $5
\times 10^3$ to $20 \times 10^3$ samples for smaller sample sizes,
$8^3$ through $64^3$, and $300$ to $5 \times 10^3$ samples for the
largest sample size, $128^3$.)  Estimates of the dimension of these
surfaces, $d_s$, can be obtained from the discrete logarithmic
derivative,
\be\label{eqn_dslocal}
\tilde{d}_s(h,L) = \ln[A(h,\sqrt{2}L)/A(h,L/\sqrt{2})]/\ln(2).
\ee
A plot of these estimates, with statistical errors, are shown in
\figref{ds_connected}.  The estimates for the case of $h=2.27\simeq
h_c$ appear to approach a fixed value, $d_s$, as $L \rightarrow \infty$, while
$\tilde{d_s} \rightarrow 2$ for $h < h_c$, as expected.  For $h>h_c$,
the apparent exponent either starts at $\tilde{d_s} > d_s$ and falls, or
first rises before dropping with $L$. This behavior presumably arises
for $L \ll \xi$, where the growing volume allows for larger surface
area, while for $L \gg \xi$, the domain walls become confined to a
distance less than $\xi$ from the right and left faces of the sample
and thus effectively become two dimensional.  From this plot and the
results for the (111) orientation (\figref{lnder_111}), we
estimate
\be
d_s = 2.30 \pm 0.04,
\ee
where systematic errors due to finite
size effects and uncertainty in $h_c$ dominate the statistical
uncertainties.

\subsection{Roughness in the ferromagnetic phase}\label{secrough}

We have verified that the surface roughness of the non-fractal
domain walls in the ferromagnetic phase
are consistent with theoretical
expectations.\cite{GrinsteinMa,HuseHenley,FisherFRG,ChauveLeDoussalWiese}
Specifically, we calculated the ``height" of the surfaces --- deviation from flat --- in anisotropic samples of
shape $X \times L^2$, with the outer two layers in the $x$-direction
fixed to be $+$ or $-$ and, as before, the sample periodic in the $y$ and
$z$ directions.
Again, to reduce lattice artifacts, we use samples whose ``$x$'' faces
are oriented in either the (100) or (111) directions.

\begin{figure}
\centering\includegraphics[width=7.cm]
{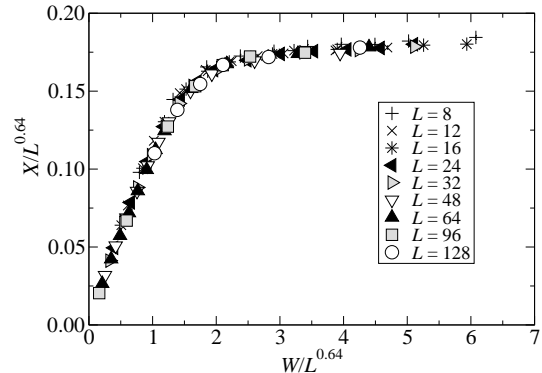}
\caption{
Scaling plot for the roughness of a forced interface in the
ferromagnetically ordered phase as a function of the aspect ratio
of an $X\times L\times L$ sample. For the values of $h$ shown here
with $h<h_c$, the width of the interface scales
as $W\sim L^\zeta$ with the best fit $\zeta = 0.64(3)$,
comparable to the expected exact result $\zeta=2/3$. The statistical
$1\sigma$ error bars
are $1/5$ of the symbol sizes or less.
}
\label{fig_scalerough}
\end{figure}

As overhangs are possible in these interfaces, it is necessary to
define  carefully  the ``height" function, $u(y,z)$,:
for a given $y$ and $z$ coordinates, we use twice the average of the
$x$-coordinates of the set of spins in ${\cal U}_{+-,--}$ ; in the absence of  overhangs, this gives
the desired surface height.
The sample averaged rms {\it  width} $W$ is defined  by
$W^2 = \overline{{[u^2]-[u]^2}}$, where the square brackets indicate the
average of $u(y,z)$ over the $y-z$ coordinates of the sample. Simple scaling in the ordered phase 
suggests that
\be
W = L^\zeta T(h,X/L^{\zeta}),
\ee
for large values of $X$ and $L$, with $T$ a geometry dependent function.
We find that using $\zeta=0.64\pm0.03$, consistent with
the expected value\cite{GrinsteinMa}
$\zeta=2/3$, describes the data fairly
well, as seen in \figref{fig_scalerough}.

\begin{figure}
\centering\includegraphics[width=8.cm]
{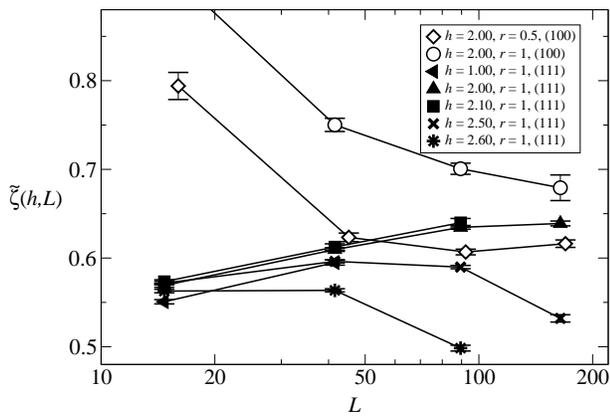}
\caption{
Plot of the effective roughness exponent $\tilde{\zeta}(h,L)$
in the ordered phase, for $h=2.0$ with
(100) oriented faces, and $h=1.0,2.0,2.1,2.4$, with (111) oriented faces. The
samples have $X=rL^{2/3}$ layers (of area $L^2$ for (100) and
area $\sqrt{3}L^2$ for (111).)
For all samples in the ordered phase,
the exponent approaches $\zeta = 0.66 \pm 0.03$ as
$L \rightarrow \infty$, consistent with the expected $\zeta=2/3$.
For comparison, data for the disordered phase is included; the apparent
exponent decreases for large systems when $h > h_c$.
}
\label{fig_localrough}
\end{figure}

The convergence of the roughness of the interface to its asymptotic form
is made more apparent by defining an effective scale dependent roughness
exponent
\be
\tilde{\zeta}(h,(L_1L_2)^{1/2}) = \ln(W_2/W_1) / \ln(L_2/L_1)
\ee
where the $X_{1,2}$ are chosen to have the values $rL_{1,2}^{2/3}$,with
$r$  fixed at close to unity.
Assuming that $\zeta$ is indeed {\em near} $2/3$, this choice ensures
that a typical
wall is found, rather than the best of a set of $\sim L^{1-\zeta}$
possibilities
that would result from using a sample that was much longer
than $L^\zeta$ in the $x$ direction.
Such a sample shape would result in the same asymptotic value for $\zeta$,
but would have (probably logarithmic) corrections to scaling.
As can be seen in \figref{fig_localrough}, the effective exponent appears to
converge to $\zeta = 0.66 \pm 0.03$ in both geometries.
Note that even with the appropriate anisotropic scaling, the corrections to scaling
are large for samples up to $L=100$  with the corresponding $X \sim 20$.

\subsection{Incongruence box-counting interface exponent $d_I$}
\label{secboxcount}

For an alternative measurement of the dimension of the domain walls at
criticality, we have used a {\it box counting} method.  In this method, we
compare the configuration given $+-$ (or $-+$) boundary conditions
with {\em both} $++$ and $--$ configurations. This is done at various
scales $w$, by partitioning the sample into $(L/B)^3$ cubes of volume
$B^3$.  If the configuration with twisted boundary conditions differs
from {\it both} $++$ and $--$ in a given volume $B^3$, that cube must
intersect the domain wall.  But this wall will {\it not} include any
boundary of frozen regions that is isolated from other broken bonds by a
distance of at least $B$.  In particular,  when $\Sigma = 0$, the number of such
intersecting boxes $N(B, L, h)$ will be {\em zero} for $B$ smaller than the
size of the frozen interior region. For example,
the $+-$ and $-+$ configurations in \figref{fig_walls_pic_2} are {\em
locally} congruent everywhere with either the $--$ or the $++$
configuration.
Thus only for boxes larger than the width
of the interior region will a domain wall be apparent.

\begin{figure}
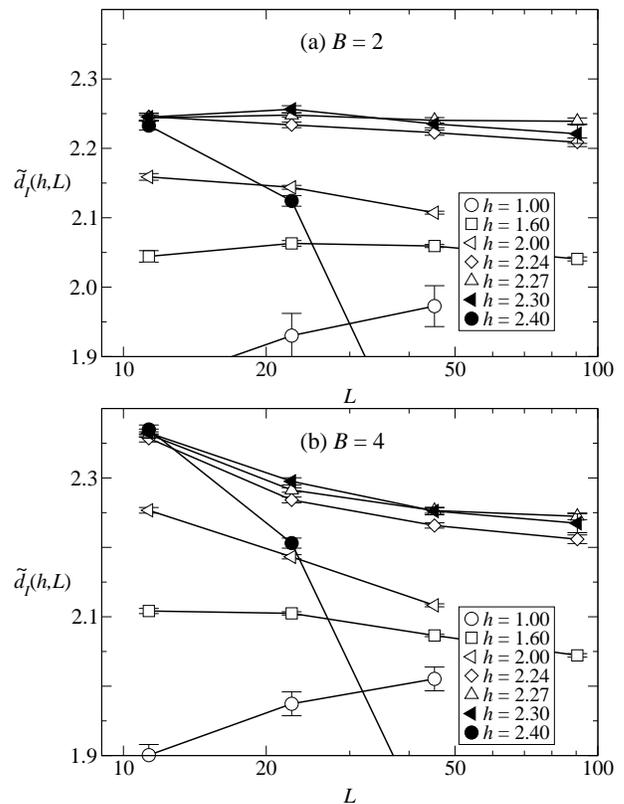

\centering\includegraphics[width=8.cm]
{15a.eps}
\centering\includegraphics[width=8.cm]
{15b.eps}
\caption{ Estimates $\tilde{d}_I(h,L)$ of the box-counting fractal
dimension $d_I=\tilde{d}_I(h_c,\infty)$, for the (100) orientation of
controlled faces.  Comparisons of the $+-$ configuration are made with
the $--$ and $++$ configurations in boxes of volume $B^3$. If the $+-$
configuration differs from both of the others, that box is considered
part of the domain wall. The finite logarithmic derivative of the
scaling of the number of such boxes with sample size $L$ yields the
estimates shown with the lines connecting data points with the same
$h$.  The error bars represent $1\sigma$ statistical uncertainties.
For $h = 2.27$, the dimension estimate converges to $d_I=2.24 \pm
0.03$, the error being a combination of statistical error and
systematic errors ($\approx 0.02$) caused by finite size effects and
uncertainties in $h_c$.  }
\label{fig_d_box}
\end{figure}

The scaling of the number of intersecting boxes $N(B, L, h)$ with $L$
gives an alternate estimate of an effective fractal dimension which we call
${d}_I(h,L)$, anticipating that $N \sim (L/B)^{d_I}$ at the critical
point (see \figref{fig_walls_define}(b).)
Using the same form of the discrete logarithmic derivative
between scales $L$ and $2L$ as in \eqref{eqn_dslocal} gives the
effective exponent $\tilde{d}_I(h,L)$, as summarized in Figs.\ \ref{fig_d_box}
and \ref{lnder_111}. This
estimate yields a constant at large $L$, within statistical
errors, for $h = 2.27 \simeq h_c$ and gives a value $d_I = 2.24 \pm
0.03$.

We note that a useful compatible definition for $d_I$
can be based on  {\it bonds}  rather than spin blocks: count the number of bonds that are broken with the $+-$ or $-+$ BCs
that are satisfied with both the $++$ and $--$ BCs. The number of 
such bonds $N'$ should have the same scaling form as $N$ does  for
fixed $B$. We have used this bond definition in a smaller number of samples and find results for $\tilde{d}_I(h,L)$
at large $L$ consistent with the spin block definition of $d_I$
defined above.

\subsection{Exchange stiffness exponent $d_J$}\label{secdJ}

A third measure that we have used to study domain wall geometry is the
contribution of the {\it exchange energy} to the stiffness $\Sigma$.
This we denote $\Sigma_J$.  It is the {\it signed sum} of the broken bond
weights, counted as negative for the $--$ and $++$ configurations and positive
for the $-+$ and $+-$ configurations.

As in computing $\Sigma$, using the symmetrized energy
differences reduces boundary effects.
If $\Sigma = 0$, then $\Sigma_J = 0$, for example, though comparing
the configurations with $++$ and $+-$ boundary conditions in such a
sample will reveal a domain wall while comparing those with $++$ and
$-+$ boundary conditions
will reveal a second entirely distinct domain wall.
Either of these domain walls, along with a portion of the frozen spins
that make up the boundary, would be counted in the method which yielded $d_s$.
But in this symmetrized
measure from $\Sigma_J$, the signed sums would cancel, so that neither domain
wall would be counted.
Similarly, when the box size $B$ is smaller than
the size of the frozen interior,
the measure used to find $d_I$ would also be zero.
Note, however, that $\Sigma_J$
{\it does} include some of the boundaries of the frozen regions but it
does so with signs that can be either positive or negative.  In the
ordered phase, then, the exchange stiffness
$\Sigma_J$ will include contributions
from the region
between the two domain walls that occur, contributions that would not
have been included in the other methods. The three proposed
measures are thus potentially all
different, especially off-critical, but perhaps also at criticality.

At the critical point the
exchange energy part of the symmetrized stiffness
will have contributions from
the domain walls with holes, $\sim L^{d_s-\beta/\nu}$, 
equivalent to the box counting
measure of the domain wall, as well as contributions from parts
of the boundaries of the frozen regions.
The simplest expectation is that the
the contributions from the frozen region
boundaries will be random in sign and thus less important {\it in
toto}.

\begin{figure}
\centering\includegraphics[width=8.cm]
{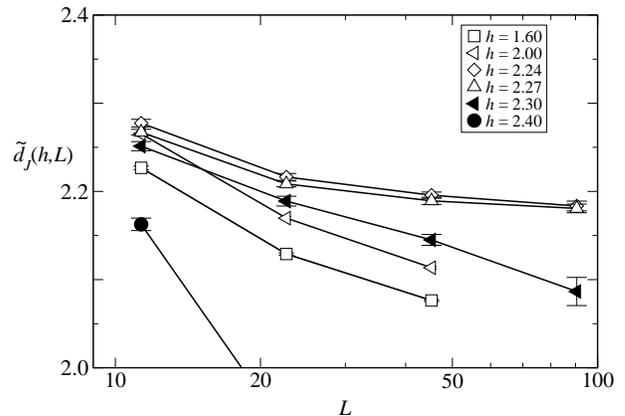}
\caption{
Estimates $\tilde{d}_J(h,L)$ of the scaling of the exchange contribution to
the stiffness defined as the total signed surface area of the
changes between
$++$, $--$, $-+$ and $+-$
boundary conditions.
The logarithmic derivative of $\overline{\Sigma}_J(L)$ gives the values shown with the lines connecting data points with the same $h$.
The error bars represent $1\sigma$ statistical uncertainties.
For $h = 2.27$, the dimension estimate converges to $d_J=2.18\pm 0.03$.
}
\label{lnder_Twistbond}
\end{figure}

\begin{figure}
\centering\includegraphics[width=8.cm]
{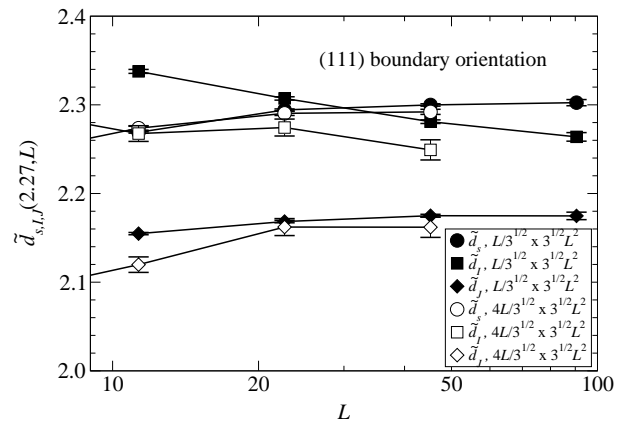}
\caption{
Estimates $\tilde{d}_s(h=2.27,L)$, $\tilde{d}_I(h=2.27,L)$, and $\tilde{d}_J(h=2.27,L)$ of the fractal
dimensions using controlled boundary surfaces
in the (111) plane of the cubic lattice which are
rhombi with sides of $L$ spins. The number of layers in the sample,
including the boundary planes, is $L$ or $4L$, as shown in the key.
The error bars represent $1\sigma$ statistical
uncertainties with the lines  connecting data points with the same $h$.
The dimension estimates converge to $d_s=2.30\pm 0.02$,
$d_I=2.25\pm 0.05$, $d_J=2.18\pm 0.03$; these are
consistent within errors with those from Figs. \ref{ds_connected},
\ref{fig_d_box}, and \ref{lnder_Twistbond}.
}
\label{lnder_111}
\end{figure}

The mean of $\Sigma_J(h,L)$ can be used to compute a
fractal-dimension-like quantity,
$d_J$, for the interface via the assumption that $\overline{\Sigma}_J \sim
L^{d_J}$ at $h_c$.  The scale dependent effective
exponents from our data at $h=2.27$, 
shown in \figref{lnder_Twistbond}, yield an estimate of $d_J =
2.18 \pm 0.02$ that appears to be slightly smaller than 
the other two dimensions $d_s$ and $d_I$.

One advantage of the exchange energy is that we can relate this
measure of the fractal dimension of the domain walls at the critical
point to the other exponents.  If a small additional exchange $\delta
J$ is added to the Hamiltonian (or equivalently if all the random
fields were decreased in magnitude by a uniform small amount) then the
change in the stiffness would be simply
\be
\delta \Sigma \approx \frac{\delta J}{J} \Sigma_J .
\ee
Since $\Sigma \sim L^\theta$ while $\Sigma_J \sim L^{d_J}$ with
$d_J>\theta$, the change in the stiffness will become of order the
stiffness itself and thus strongly modify the system when $L \sim
(\delta J)^{-1/(d_J-\theta)}$.  This crossover length is thus a measure
of the correlation length, $\xi$, and we thus expect the exponent
equality
\be 
\frac{1}{\nu}=d_J-\theta.
\label{scale_dJ}
\ee
This can be derived directly from the scaling form
\eqref{sigma-bar-scaling} by differentiating with respect to $J$
(equivalently with respect to $-h$) and noting the thermodynamic
identity between derivatives with respect to coefficients of terms in
the Hamiltonian and expectations of the corresponding term.  (Note that
this is closely analogous to the relation between $\nu$ and the
energetic part of the interfacial free energy at conventional finite
temperature critical points.)
Assuming the scaling relation
\eqref{scale_dJ} would give $\nu = 1.45\pm0.10$,
a slightly
different, but consistent, value of $\nu$ than that
from the scaling of the total symmetrized stiffness $\Sigma$.

\subsection{Comparison of domain wall exponents}

Due to the subtleties introduced by frozen islands and the
representation of the Hamiltonian as the sum of domain wall  and random field
components, there are three natural measurements of the
domain wall surface and the domain wall contributions to the stiffness.
Each measure has its own physical meaning.
We will argue in \secref{fract} that the difference
between $d_s$ and $d_I$ is due to frozen islands, and hence
$d_s-d_I$ should be related to $\beta/\nu$.

\section{magnetization}\label{mag}

Having established the location and order of the transition, we now focus
on an apparently problematic quantity: the magnetization.
Given sets of ground state spin configurations $\left\{s_i\right\}$,
the distribution of magnetizations can be studied as a function of
$L$ and $h$.
In order to better understand the large
volume limit, we have computed the magnetization distributions
for {\em five different boundary conditions}:
all boundary spins fixed to a single value, either all positive or all negative ($F$);
boundary spins fixed at independent random values ($R$);
open boundary conditions ($O$); periodic boundary conditions ($P$);
and a combination ($Q$) with conditions
$P$, $O$, and $R$ along each of the three
axes .  The fixed spin boundary conditions will tend to favor ferromagnetism, the random will tend to favor a disordered phase and the combination $Q$ appears to significantly reduce some finite size effects.

\begin{figure}
\centering\includegraphics[width=8.cm]
{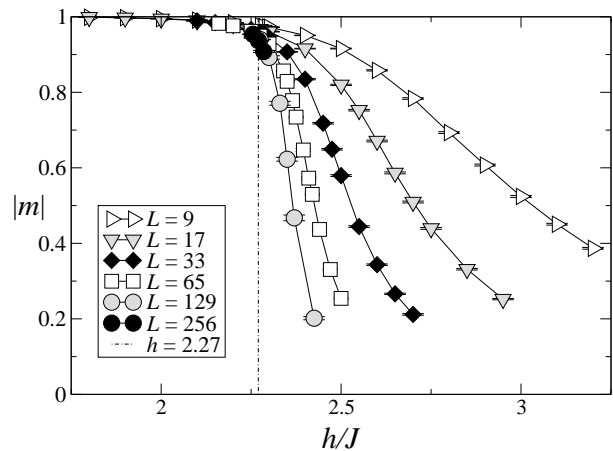}
\caption{
Mean absolute magnetization per spin, $|m|$, plotted vs.\ $h$ for various
$L$, for periodic boundary conditions.
}
\label{fig_absmag_vs_h}
\end{figure}

\begin{figure}
\centering\includegraphics[width=8.cm]
{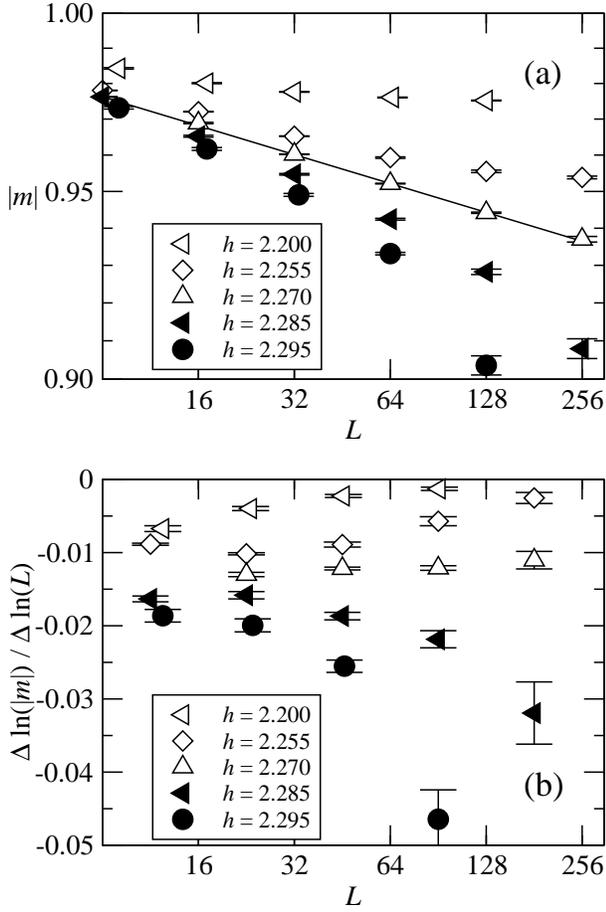}
\caption{
(a) Mean absolute magnetization per spin, $|m|$,
plotted vs.\ $L$ for various $h$
(with periodic boundary conditions.) The solid line is 
$|m| = (1.0009)L^{-0.012}$.
(b) The discrete logarithmic derivative
$\Delta \ln(|m(L)|)/\Delta\ln(L)\equiv\ln(|m(L')|/|m(L))/ln(L'/L)$,
vs.\ $\sqrt{LL'}$, with $L' \approx 2L$. This is used
to directly
estimate $\beta/\nu$, yielding
$\beta/\nu = 0.012 \pm 0.004$, where the error bars are dominated by the
range of values for $h_c$ obtained by fitting over sizes up to $L=256$.
}
\label{fig_meanabsmag}
\end{figure}

We first describe our results for the mean of the absolute value of the
magnetization density
\be
|m| = |\sum_{i} s_i|L^{-d}
\ee
for cubic samples
with periodic boundary conditions (P).
\figref{fig_absmag_vs_h} is a plot of our data as a function of $h$, for
various $L$; the magnetization drops off quite steeply near $h_c$.
\figref{fig_meanabsmag} shows the magnetization as a function
of system size, along with its discrete logarithmic derivative,
which yields an effective scale dependent exponent.
To within errors, the magnetization is consistent with
power law scaling,
\be
m \sim L^{-\beta/\nu},
\ee
with $\beta/\nu = 0.012 \pm 0.004$. For $h<h_c$, the magnetization appears
to approach a constant (e.g., $m(2.255,L\rightarrow\infty) \approx 0.952$).
For $h>h_c$, the effective exponent decreases significantly
as $L$ increases.

\begin{figure}
\centering\includegraphics[width=5.3cm]
{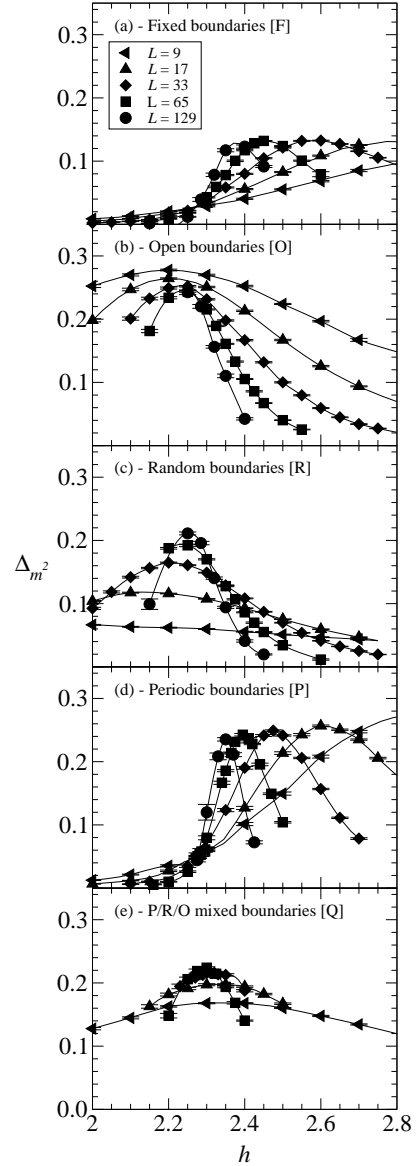}
\caption{
Magnitude of sample-to-sample fluctuations $\Delta_{m^2}$
in the mean square magnetization per spin, as a function of random field strength $h$,
for various system sizes, $L$,
(a) for fixed $s_i=+1$ boundary conditions ($F$),
(b) for open (free) boundary conditions ($O$),
(c) for random fixed spin boundary conditions ($R$),
(d) for periodic boundary conditions ($P$),
and (e) for mixed periodic, random fixed,
and fixed boundary conditions, one along
each axis ($Q$).
The curves 
are fit locally with Gaussians
in the regime where $\Delta_{m^2}$ is greater than approximately
$3/4$ of its peak value.
Extrapolating the peak locations
to $L=\infty$ gives a best fit value of $\nu = 1.38 \pm 0.08$
and $h_c = 2.272 \pm 0.004$, with the dominant errors being
systematic errors arising from variations
in the extrapolated values,
presumably due to corrections to scaling.
The lines shown are spline fits to visually organize the data.
}\label{fig_delta-m2}
\end{figure}

For further
analysis, we characterize the distributions of $m$
by the average over samples
of the square of the magnetization per spin, $\overline{m^2}$,
and the root-mean-square sample-to-sample variations of the square of the magnetization
\be 
\Delta_{m^2} \equiv \sqrt{\overline{m^4}-(\overline{m^2})^2}  .
\ee 
Our results for $\Delta_{m^2}$ are shown in
\figref{fig_delta-m2}.
As $L$ is increased,
the peak magnitude of $\Delta_{m^2}$ is seen to decrease for some boundary conditions
$F$, $O$ and $P$, while it
increases for others $R$ and $Q$.
For boundary conditions $R$, $O$, $P$, and $Q$, the peak
heights appear to be converging to a similar
fixed value, bracketed from above and
below by the different sets of data.
In general, we would expect that the height of these peaks would scale
for asymptotically large sizes as $L^{-2\beta/\nu}$; the data are thus
consistent with either $\beta=0$ or with a very small $\beta$.

One can estimate the location of the transition by fitting the data
for $\Delta_{m^2}$ near the peaks at five or more values of $h$ to a
Gaussian form. (The Gaussian gives a better fit than a parabolic form
over a larger range of $h$, though either form should give the same
limit for $h$ near enough to the peak and $L$ large enough.)  The fitted
location of the peaks is extrapolated for all boundary conditions as a
function of $L$.  We obtain agreement of the extrapolations for
$1.3 < \nu < 1.45$ with a value of
$h_c = 2.272 \pm 0.004$, consistent with the value from $P_0$ and other
estimates.
We believe that this independent
estimate is relatively precise and robust, due to
the variety of boundary conditions used, with the variation in the
results giving an estimate of systematic uncertainties.

For the fixed spin boundary conditions, $F$, the peak magnitude of
$m^2$ is apparently converging to a {\it different} value (note that
the magnetization near the surfaces will vary less than with the other
boundary conditions.) If either these data, $F$, or the periodic
boundary condition data, $P$, {\it at the critical point} are used
(rather than the data near the peak) then a smaller value of
$\Delta_{m^2}$ is found, roughly the same although apparently still
distinct for these two cases.

Collectively, our magnetization data would appear to suggest a picture
of the transition that is consistent with that of reference
Ref.\onlinecite{MachtaNewmanChayes}:
three possible ``states" at the critical
point, ``up", ``down", and ``disordered" as would occur at a first
order para- to ferro- magnetic transition.  As we shall see, however,
our other data and further thought suggest that this picture, while a
very good approximation, is not correct.  We will argue that in fact
$\beta$ is small but non-zero and thus in astronomically large samples
the magnetization will decay slowly to zero at the critical point but
with the scaling functions for the distribution of the magnetization
(and their moments) depending on the type of boundary conditions as is
the case for pure systems at conventional critical
points.\cite{BC-pure-mag} Data suggesting this is
presented in the next section.

\section{Spin clusters  at criticality}\label{clust}

The distribution of the magnetization studied above gives some
information about the ground state correlations of the RFIM.  But
because ground state correlations between Ising spins are controlled
by the probability that a pair of spins of interest are in opposite
directions, the observation that the magnetization at the critical
point tends to be rather close to saturation suggests that the loss
of correlations as the random field is increased through the critical
point may be associated with rather rare events.  In this section, we
investigate the nature of the effect that we believe gives the dominant
contribution:
the occurrence of connected clusters of spins of the one sign
completely surrounded by spins of the opposite sign. Because all of
the exchanges are ferromagnetic, such {\em isolated inverted clusters} will,
{\it a fortiori}, not change when the boundary conditions are
inverted: either the spins surrounding them will flip in which case
they will be content the way they were, or the surrounding spins will
not flip and the spins in the cluster will be isolated from the
boundary condition change. Thus these isolated spin clusters are
frozen.

\begin{figure}
\centering\includegraphics[width=8.cm]
{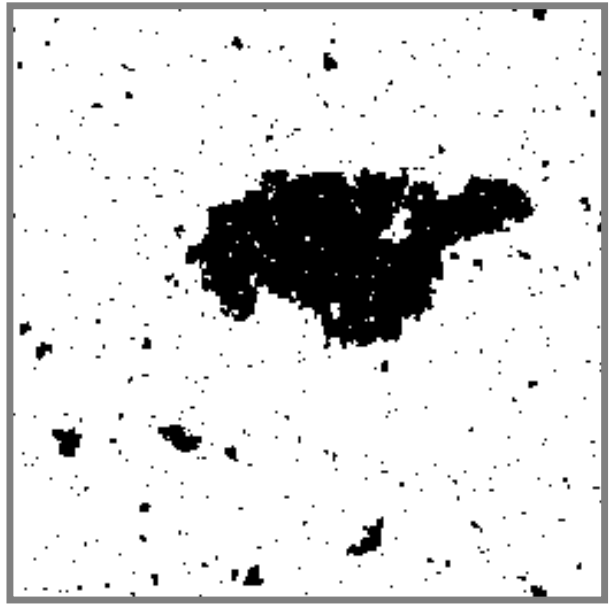}
\caption{
A slice of a spin configuration in a $256^3$ sample at $h=2.27$.
The dark squares indicate an up spin. The nesting of spin clusters
can be seen here --- the number of levels of nesting, the ``depth'',
of the full configuration is $k=3$.
The domain walls are determined by working recursively
inwards from the majority
(down) spin cluster. The surface of each cluster is taken to be the
outer surface and does not include the surface of subclusters.
}
\label{fig_slice}
\end{figure}

We have computed the statistics of the domain walls
that enclose isolated spin clusters in $5 000$ or more samples of system sizes up to $128^3$ and $1000$ samples
of size $256^3$ at $h=2.27\approx h_c$.  A slice of a configuration is
shown in \figref{fig_slice}. Statistical errors in the
dimension estimates and number distribution were computed by a bootstrap method
(resampling the statistics over the computed configurations);\cite{ShaoTu}
the error bars indicate the estimated rms fluctuations in the statistics
at each cluster size.

We note that previous work by Esser, Nowak
and Usadel\cite{EsserNowakUsadel}
studied the domain structure for a single sample size. They address
questions of percolation in the 3D Gaussian RFIM, but they claim that
the cluster distribution is not broad. We find, in contrast, that there
is a broad tail, which, though weak for smaller systems, becomes more
important as $L$ increases at the critical point. To directly contrast with the
results of \refref{EsserNowakUsadel}, we find that
the sum of the volume fraction of the {\em two} largest clusters,
though near $1$, slowly
{\em decreases} as $L$ increases, at $h=2.27\approx h_c$.
The transition separates a state with one
infinite connected set of spins of the same sign
from a disordered state with two antiparallel
incipient infinite clusters.

\subsection{Cluster surface}\label{secsurf}

\begin{figure}
\centering\includegraphics[width=8.cm]
{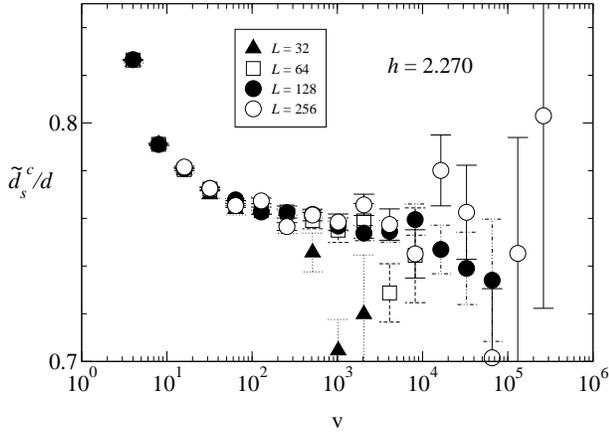}
\caption{
Dependence of the surface area (number of broken bonds)
of cluster boundaries on the enclosed
volume, $v$, expressed as an effective exponent $\tilde{d}_s^c(h,L,v)$, at
$h=2.270\approx h_c$, for $L=32,64,128,256$. The cluster surface area
scales as $v^{0.755\pm0.007}$ for the largest clusters that are not affected
by finite size effects,
yielding a fractal
dimension  $d_s^c = 2.27(2)$ for the cluster surfaces.
}
\label{fig_clustersurface}
\end{figure}

For each cluster, the total volume, $v$, --- which includes the volume
of ``holes" of opposite spin --- was computed, as was the surface
area, $a$, of the cluster: the number of unsatisfied nearest neighbor
bonds that separate the cluster from its {\em surrounding} region of
opposite spin. The domain walls are found recursively, taking as the
initial surrounding region the majority spin cluster, which typically
occupies $>97\%$ of the volume at $h_c$ for $L<256$.\cite{depthnote}
Binning the clusters by volume $v$, logarithmically spaced by powers
of 2, averaging the surface area in each bin, and taking the discrete
logarithmic derivative gives an estimate of the fractal dimension of
the cluster surfaces, dimension $\tilde{d}_s^c(h,L,v)$. As indicated
in \figref{fig_clustersurface}, at the critical point the surface area
appears to scale as $v^{0.755\pm0.07}$ for intermediate size clusters
with $1\ll v\ll L^3$.
The error in this exponent includes both statistical error and
the apparent uncertainty of corrections to scaling that are affecting
the convergence to a constant value. This value is little
affected by the estimate of the location of $h_c$ (varying $h$ changes
the number of clusters, but, within the uncertainty of $h_c$, does not
affect the geometry of the domain walls.)
We have verified that the volume enclosed by the domain walls separating
opposite spins scales in a manner numerically consistent with this
{\em volume} being {\em nonfractal:}
\be
v \sim \ell^{d},
\ee
with $\ell$ being either the
geometric mean or the maximum of the lengths of the sides of the minimal rectilinear
box that encloses the cluster.
The
extrapolation of $d_s^c(h,L,v)$ to large $L$ and $\ell$ is therefore consistent with
clusters having typical diameter $\ell\sim v^{1/d}$ and typical
surface area 
\be
a\sim \ell^{d_s^c}
\ee
with
\be
d_s^c \simeq 2.27 \pm 0.02
\ee
a fractal surface dimension consistent within the statistical
uncertainties with our estimates of the fractal dimensions $d_s$ and $d_I$ of
the domain walls induced by changing boundary conditions at the
critical point. In particular, this surface dimension bears a close
resemblance to the dimension of the spanning surface which we denoted
$d_s$; thus we conjecture that
\be
d_s^c=d_s.\label{eqnds}
\ee

\subsection{Cluster density}\label{secrho}

\begin{figure}
\centering\includegraphics[width=8.cm]
{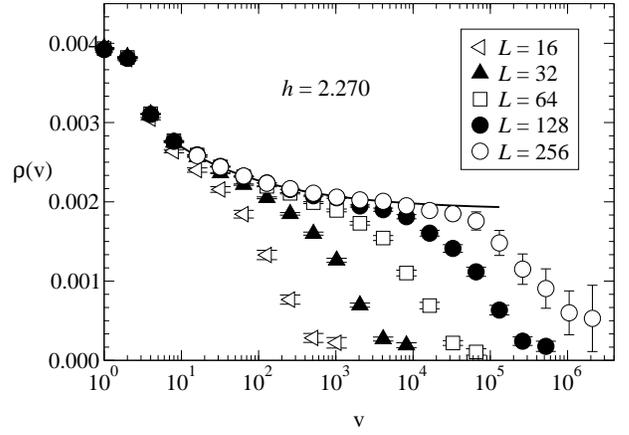}
\caption{
Fraction of the volume $\rho(v)$ occupied by clusters of volume
between $v$ and $ev$, at $h=2.270\approx h_c$,
found by normalizing the data binned according
to powers of $2$ (i.e., dividing the volume fractions in
the $[v,2v]$ bins by $\ln 2$.)
The solid line is 
$\rho(v) = 0.0019+(0.0017)v^{-0.33}$, one of the trial fits used to
extrapolate to large $v$.
}
\label{fig_clustercount}
\end{figure}

\begin{figure}
\centering\includegraphics[width=8.cm]
{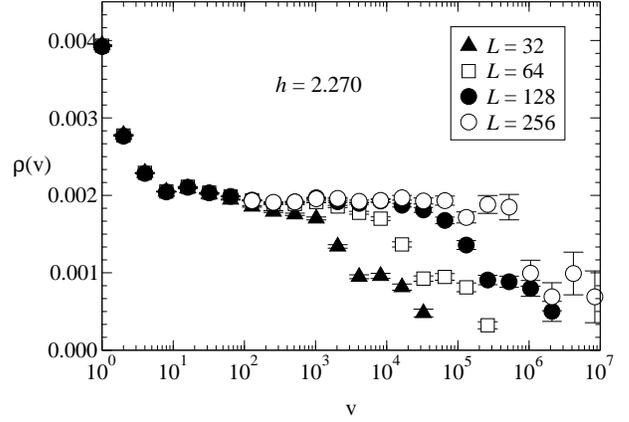}
\caption{
Fraction of the volume $\rho(v)$ occupied by clusters that
are contained in {\em rectilinear volumes} (``boxes'')
between $v$ and $ev$, at $h=2.270\approx h_c$,
found by normalizing the data binned according
to powers of $2$ (i.e., dividing the volume fractions in
the $[v,2v]$ bins by $\ln 2$.)
For large $v$, $\rho(v)\rightarrow 0.0019 \pm 0.0002$, if $h_c=2.270$.
}
\label{fig_clustercount2}
\end{figure}

\begin{figure}
\centering\includegraphics[width=8.cm]
{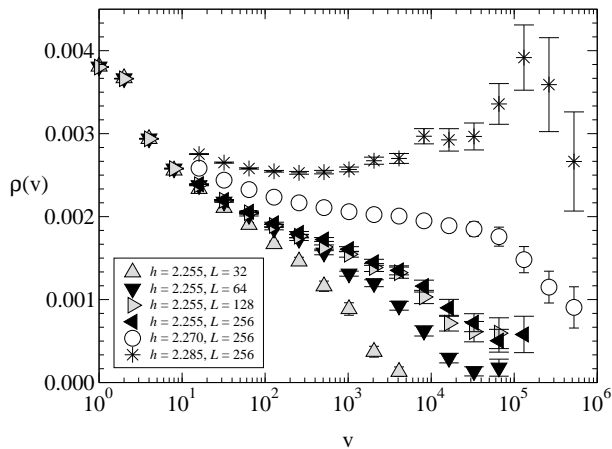}
\caption{Fraction of the volume $\rho(v)$ occupied by clusters of
volume between $v$ and $ev$, found by normalizing as in
\figref{fig_clustercount}.  Here, the volume fractions are plotted for
$L=32,64,128$, with $h=2.255$, and $L=256$, with $h=2.255$, $2.270$
and $2.285$, to indicate some of the effect of changing $L$ or $h$ on
$\rho(v)$.  The data for $h=2.255$ apparently converges at large $L$
to a well defined distribution that has a finite-$v$ cutoff,
consistent with a finite correlation length in the ordered phase.
This is in contrast with the data for $h=2.27$, the putative critical
point, for $L\le 256$, as seen in \figref{fig_clustercount}.  For
$h=2.285$, in the disordered phase, large volume clusters start to
occupy a larger fraction of the volume than smaller clusters, for
$L=256$.}
\label{fig_clustercount_cf_h2}
\end{figure}

New information is given by the {\it densities} of the clusters as a
function of their size, in particular their dimensionless volume
fraction
\be\label{eqcluster}
\rho(v)\equiv \frac{v}{\delta v}\pr[{\rm site} \in {\rm
cluster \ of\ size \ in}\ (v,v+\delta v)] . 
\ee
From the data in \figref{fig_clustercount} (and for the slightly
different measure of \figref{fig_clustercount2}, where $v$ is the
volume of the smallest parallelpiped of fixed orientation
enclosing the cluster) we see that
clusters that are neither too small nor limited by finite size effects
--- roughly a decade in length scale for $L=256$ --- occupy an
approximately scale independent volume fraction.
A comparison of the cluster distributions for nominally off-critical
values of $h$, as seen in \figref{fig_clustercount_cf_h2}, shows how
$\rho(v)$ depends on $h$.
From these plots we infer a large-$v$
limit of
\be
\rho(v) \rightarrow \rho_\infty \simeq 0.0019 \pm 0.0004.
\ee
We cannot, of course, rule out a slow decrease of $\rho(v)$ to zero
for large volumes, especially as our effective range of length scales
here is less than for other quantities because of the restrictions due
to finite size effects.  But we {\it can} understand on the basis of
our other observations why one should expect a small but non-zero
value for $\rho_\infty$.

\section{Scaling}\label{scale}

In this section we pull together our various results about domain
walls, stiffness, magnetization, and inverted spin clusters and show
how they are all consistent with a simple picture of scaling
behavior at a zero temperature phase transition.

\subsection{Critical correlations}\label{seccritcorr}

At the critical point, the energy cost of domain walls is typically
sufficiently large that almost all cubical samples --- about $96\%$ of
them --- would rather have no spanning (or other large scale) domain
walls unless forced to by boundary conditions. But in a small fraction
of the cubical samples the random fields in the central region are
sufficiently strong that they force the system to have {\it two}
domain walls for one of the two ``ferromagnetic" ($++$ or $--$)
choices of boundary conditions.  In samples that are twice as long,
this occurs much more frequently as evidenced by the increase,
on going from cubical
to elongated samples,  in the
probability, $P_0$, that the stiffness vanishes.  Although
whether such a pair of walls is
favorable generally depends on both the random fields in the whole
system and the local behavior near the walls, a crude picture of what
is going on can be drawn by assuming that the wall energies are
relatively local and weakly dependent on each other.  We restrict
consideration for now to the critical point.

First consider a system of dimensions $\frac{1}{2}L\times
L\times L$ with the boundary conditions imposed on the faces
perpendicular to the short axis. Assume that the
probability that the single wall
energy, $E_{+W}^\ell\equiv E_{+-}-E_{++}$, in such a
system is {\it negative} is $q\ll
1$.  Crudely, for two walls to be favorable in a cubic system with
$++$ boundary conditions, as is needed to make $\Sigma=0$ , one must
have {\it both} $E_W^\ell<0$ for the left half of the system and $E_W^r \equiv
E_{-+}-E_{++}<0$ for the right half of the system.
Naively, this occurs with probability of order $q^2$. (More
precisely, one of the $E_W$'s could be positive but not by enough to
dominate the other one; this will not change things much as long as the bulk of the distribution of the
$E_W$'s is skewed substantially to the positive side of zero.)
 But in a system
of length $2L$ rather than $L$, there are many more possibilities: if
we divide the system into four sections of length $L/2$, one
could have, for example, the second from the left having $E_W^\ell <0$
and the rightmost having $E_W^r<0$ with the wall energies of the other sections being positive.  As there are six such choices
among the four sections of the elongated system, we expect that the
chances of having $\Sigma=0$ will be about six times as large as in
the cubical system --- obviously a very crude approximation, but one
that yields roughly the measured magnitude of the ratio $P_0(2L\times L\times L)/P_0(L\times L\times L)$. Note that this
picture implies that for systems that are much longer than they are
wide, the typical number of domain walls in the ground state will grow
linearly with the length. The roughly random spacing between them will
lead to exponential decay of the end-to-end correlations in such a
system, with a characteristic length proportional to the linear dimension of
the cross-section as should be expected on general finite size
scaling grounds.

At conventional critical points in two dimensions, conformal
invariance relates the exponential decay of correlations in long tubes
to the power law decay of correlations in the bulk in infinite
systems: the exponent $\eta$ is simply proportional to the ratio of
the width-dependent correlation length to the
width.\cite{conf-invariance} In our case, there is no such exact
relation, but one can make a qualitative argument that suggests a
similar result.
Consider a region of diameter of order $\ell$ centered on some
chosen spin in the bulk of the sample and assume that outside of this
region, the spins in the vicinity are $+1$. The only way that the
spins inside the region of interest can be $-1$ is if there is a
domain wall relative to the pure ``up" configuration which surrounds
this region and has negative energy.  Roughly speaking, such a closed
domain wall must be made up of four or more sections which are joined
together with each having negative (or close to zero) energy.  Since
the amount of freedom perpendicular to the area of each of these will
be somewhat less than their linear dimensions, a crude approximation
is that the probability of finding each such section is of order $q$
and the probability of finding the total domain wall energy negative
is of order $q^4$.  With $P_0[{\rm cube}]\simeq 0.04 \sim q^2$, this
suggests that the probability for finding such a spin flipped region
will be of order $\rho_\infty \sim q^4 \sim 0.002$, in the same range
as that found.  Although the above argument is very crude --- factors
of two's or pi's could easily have been fudged in! --- it nevertheless
provides a suggestive connection between the smallness of various
quantities.  Indeed, it actually provides more: a precise method for
estimating $\beta/\nu$.

Consider a single spin in the center of a large system at the critical
point with, for simplicity, $+$ boundary conditions . For each factor
of $e^{1/d}$ in length scale, $\ell$, there is a probability
$\rho_\infty$ that the spin is an element of a cluster, flipped with
respect to its surroundings, with volume in the associated range $\ell^d
\le v< e \ell^d$. There will, of course, be correlations between the
probabilities of occurrences of such inverted regions that are similar
in size and nearby to one another. But, because inverted regions are so
rare, the effects of these correlations will be negligible and we can
assume that each range of $\ell$ around the chosen spin is
independent.
A simple picture of the behavior then emerges: the spin of
interest will be in a cluster of one orientation of diameter $\ell_1$,
which itself will be in a much larger cluster of the opposite
orientation of size $\ell_2$, etc. with the successive sizes growing
approximately geometrically --- more precisely as a Poisson process in
$\ln(\ell)$ with density $ d\rho_\infty $ (with $d=3$).  The probability,
$p_\parallel$, that the spin has the same orientation as the largest
cluster --- the system size $L$ --- can readily be computed from the
properties of the Poisson process; this result yields the mean value
of the spin given fixed ($+$) boundary conditions

\be
\overline{s}^{(+)}= 2p_\parallel -1 \sim \frac{1}{L^{\beta/\nu}}
\ee
with the exponent
\be
\frac{\beta}{\nu}\approx2d \rho_\infty  \simeq 0.011 \pm 0.003
\ee 
A similar argument for two spins a distance $|x-y|$ apart in an infinite system gives 
\be
\overline{s_x s_{y}} \sim \frac{1}{|x-y|^{d-2+\tilde{\eta}}} \label{zero-correlations}
\ee
with the modified ``anomalous dimension"  exponent for these {\it untruncated} zero-temperature correlations given by
\be
d-2+\tilde{\eta}= 2\beta/\nu \approx 4d\rho_\infty .
\ee

This picture of droplets within droplets strongly suggests that
{\em neither} spin species will percolate at the critical point.  This is, a
priori, rather surprising as the percolation concentration for a three
dimensional cubic lattice is substantially less than half and so one
might have expected both spin species to percolate even somewhat into
the ordered phase.  The fact that they do not in this system is
associated with the smallness of $\beta$ and the nature of the
critical point.

In practice, unfortunately, the value of $\beta/\nu$ is so small for
the 3D RFIM that the effects  discussed above will be all-but
unobservable even if experiments could reach equilibrium.  But in
higher dimensions, four or five, they might be observable numerically
as relatively large systems sizes (e.g., more than $32^4$)
can be explored.

\subsection{Fractal dimensions of domain walls}\label{fract}

The picture developed above suggests that the various fractal
dimensions of interfaces or domain walls at the critical point will
{\it not} be the same but might, nevertheless, be related to the other
exponents.  The fractal dimension of the spanning interface $d_s$  (and
the dimension of the surfaces of clusters) is the dimension of a true
surface, one with no holes in it. Such a surface cuts across the whole
system but  the sets of sites it is separating cannot really be
thought of as belonging to different states --- the ``up" and ``down"
states --- since, in an asymptotically large system, most of the sites
will be frozen and unaffected by the boundary conditions.  In
contrast, the incongruence box counting dimension, $d_I$, is
sensitive only to those parts of the system that {\it are} affected by
boundary conditions: a fraction of order $1/L^{\beta/\nu}$.  A natural
conjecture is that the box counting dimension is the same as that of
the intersection of a typical fractal spanning surface with a fractal
set of dimension $d-\beta/\nu$ yielding:
\be
d_I=d_s-\beta/\nu .
\ee
This picture is somewhat analogous to what would occur right at
percolation in a diluted ferromagnetic Ising system at zero
temperature: in a finite fraction of the samples, forcing a domain
wall by changing the spin boundary conditions would cost no energy,
while in the rest it would cost an energy proportional to  the area of an
interface that only cuts across the fractal incipient infinite
cluster; this interface would have dimension analogous to our $d_I$.

The exchange energy dimension, $d_J$ is sensitive to both the frozen and the unfrozen regions.  
But a reasonable guess is that this is dominated by the unfrozen regions as the frozen 
regions contribute random signs. This suggests that 
\be 
d_J=d_I=\theta+1/\nu
\ee 
which, if correct, implies that a relation obtains between $d_s$ and the other exponents :
\be
d_s=\theta+ \frac{1+\beta}{\nu}.\label{eqnds2}
\ee

We should note, however, that these conjectures are difficult to test
in three dimensions, due to the smallness of $\beta$.
Our estimated exponents are in slight disagreement with these
conjectures, but corrections
to scaling that are not apparent
can be important at this level of accuracy. Nevertheless, 
our apparent values for $d_s$ and $d_s^c$ do appear to be larger
than $d_I$ and $d_J$.

In higher dimensions, testing the conjectured
scaling relations between these dimensions and the
other exponents might be feasible. It is of course
possible, however, that further analytic understanding would imply
that at least some of the fractal dimensions could be independent
exponents.

\subsection{Specific heat}

The specific heat of the RFIM can be experimentally
measured\cite{JaccarinoBiref} and is of theoretical importance.
Monte Carlo methods at finite temperature have been used to estimate
its value.\cite{RiegerRFIM,RiegerYoung}
In addition,
Hartmann and Young (HY) have recently\cite{HartmannYoung}
computed the exponent $\alpha$ describing
the divergence of the specific heat, using
ground state configurations. They
find a value of $\alpha = -0.63 \pm 0.07$.
Using the {\em same} thermodynamic
assumptions, but different analysis
methods, we find
$\alpha = -0.01 \pm 0.09$.

One expects\cite{HartmannYoung} that the
finite temperature definition of the specific heat can be extended to
zero temperature, with the second derivative of $\left<E\right>$
with respect to
temperature being replaced by the second derivative of the
ground state energy density $E_{\rm gs}$ with respect
to $h$, or equivalently up to constants, $J$. The first derivative
$\partial E_{\rm gs}/\partial J$
is just the average number of unsatisfied bonds per
spin, $E_J = L^{-d}\sum_{<ij>} s_i s_j$. Hartmann and Young (HY)
calculate the needed
second derivative by finite differences of $E_J(h)$ for
values of $h$ near $h_c$.
($E_J$ is not explicitly dependent on $J$, but changes discontinuously
in a finite sample when the spin configuration $\left\{s_i\right\}$ changes;
the second derivative is thus a set of
$\delta$-functions which are smoothed by the finite differencing.)
The finite-size scaling form assumed is that
the singular part of the specific heat $C_s$ behaves as
\be
C_s \sim L^{\alpha/\nu}\tilde{C}[(h-h_c)L^{1/\nu}].
\label{Cscale}
\ee
HY determine $\alpha$ by fitting to the maximum of the
peaks in $C_s$, which occur at $h_{\rm peak}(L) -h_c \sim L^{1/\nu}$.

\begin{figure}
\centering\includegraphics[width=8.cm]
{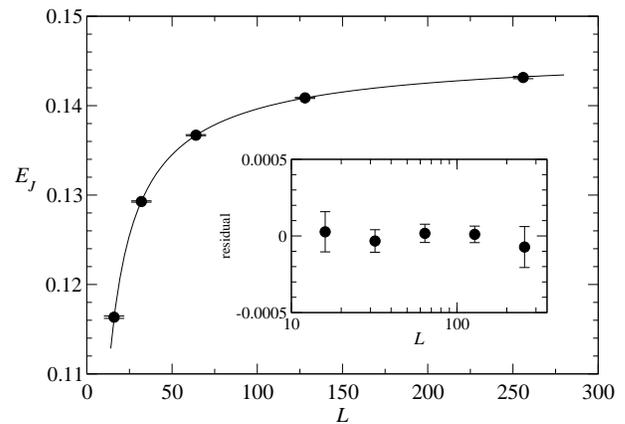}
\caption{
A plot of $E_J$, the bond part of the energy density, for
$h=2.27$, as a function
of $L$.
The fit shown is of the form $E_J = c_1 - c_2 L^{(\alpha-1)/\nu}$,
with $c_1=0.14632$, $c_2=0.29098$, and $(\alpha-1)/\nu = -0.82$.
The residuals (inset) give $\chi^2 = 0.65$. The statistical
error in $(\alpha -1)/\nu$ for fixed $h$ near $h_c$
is $0.02$, but the uncertainty
in this ratio is $0.10$ due to the uncertainty in $h_c$.
}
\label{fig_alpha}
\end{figure}

\begin{figure}
\centering\includegraphics[width=8.cm]
{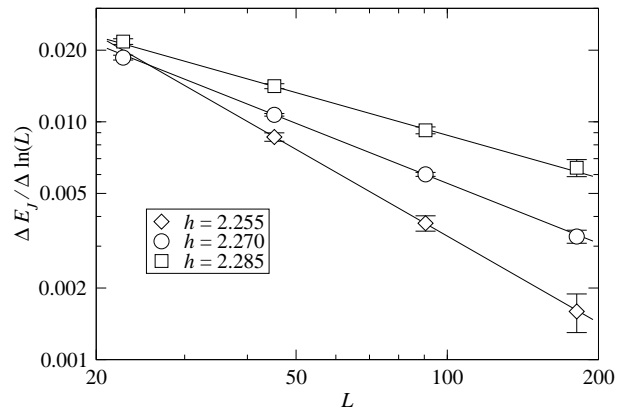}
\caption{
Plots of the discrete derivative with
respect to $\ln(L)$ of $E_J$, for $h=2.255$, $2.270$, and $2.285$.
The solid lines show power law
fits for $L\ge 30$, with slopes $1.21$, $0.84$,
and $0.60$, respectively. Using the error estimate for $h_c$,
this gives $(\alpha-1)/\nu = -0.84 \pm 0.10$, consistent with the
results from plots as in \figref{fig_alpha}.
}
\label{fig_derEJ}
\end{figure}

Here, we estimate $\alpha$ using the
results for the stiffness
from \secref{sas} 
and also by
studying the behavior of $\overline{E}_J$ {\em at} $h_c$.
The first estimate found by applying \eqref{hyper}
with our values of $\theta$ and
$\nu$, is $\alpha_{(1)} = -0.07 \pm 0.17$.
The computation from the behavior of $\overline{E}_J$ is based on
integrating \eqref{Cscale} up to $h_c$, which gives the dependence
\be
\overline{E}_{J,s}(L,h=h_c) = c_1 + c_2 L^{(\alpha-1)/\nu},
\label{EJform}\ee
with $c_1$ and $c_2$ constants.
We have computed $E_J$ for a large number of samples of various
sizes and estimated the singular part of the sample average.
We directly fit our data for $\overline{E}_J(L)$, at fixed $h$,
to the form \eqref{EJform}.
The fit for the nominal $h_c$, $h=2.27$, is shown in
\figref{fig_alpha}. The fitted values are
$(\alpha-1)/\nu = -0.82 \pm 0.02$,
{\em where the quoted error is purely statistical}. The fit is good for
$16\le L \le 256$, with $\chi^2 = 0.65$ for a three parameter fit
to five data points.
This fit is also consistent with that found
from taking the derivative of $\overline{E}_J$ with
respect to $\ln(L)$,
\be
\frac{d\overline{E}_J}{d(\ln\,L)} \sim L^{(\alpha - 1)/\nu},
\ee
at $h=2.27$,
which removes the need to fit to $c_1$, but introduces larger
uncertainties, due to the derivatives.
This data for $h$ near $h_c$ is displayed in \figref{fig_derEJ}.
By varying $h$ ($h=2.255$, $2.280$),
we estimate the systematic errors, given our
uncertainty in $h_c$, arriving at the
value
$(\alpha-1)/\nu = -0.82 \pm 0.10$,
which, using $\nu = 1.37 \pm 0.09$, gives a second estimate
$\alpha_{(2)} = -0.12 \pm 0.16$.

Besides the uncertainties in $h_c$, this result for $\alpha$ is
affected by finite size corrections. We now argue that these
corrections can be reduced by extrapolation and that a connection
exists between $\alpha$, $d_s^c$ and $\beta/\nu$:
$E_J$, being the bond part
of the energy density, is simply given by the density of domain walls,
whose scaling can be found from the results in \secref{clust}.
Namely, taking the surface area of clusters to scale with linear
size $\ell$ as $A \sim \ell^{d_s^c} \sim v^{d_s^c/d}$
and using the constant limit for the
distribution of volumes $\rho(v) d[\ln(v)]$ at large $v$, the
domain wall density in a finite sample is found by integrating
the wall density, taking into account intersections
between the scales, over $L$ up to the system size, giving
\be
\overline{E}_{J,s} \sim L^{d_s^c - d - \beta/\nu}.
\label{clustEJ}
\ee
This exponent can be justified by considering the change in
$\overline{E}_{J,s}$ upon doubling the system size.
With finite probability ($\rho_\infty \ln 2$), an extra domain wall
of scale $L$ will be introduced. The connected surface of the
domain wall will have area $L^{d_s^c}$, but the
{\em increase in domain wall area}
will be smaller, as the domain wall will intersect frozen regions.
The fraction of the sample that is not frozen scales as $L^{-\beta/\nu}$ at
criticality; the intersection of the new wall and the unfrozen region
therefore
scales as $\sim L^{d_s^c - \beta/\nu}$, so that the expected
fraction of newly
broken bonds (compared with the smaller sample)
is $\sim L^{d_s^c - d - \beta/\nu}$. (The domain wall intersects
frozen regions that did not have surfaces, as they were embedded in
like spins, adding total area, and also
intersects frozen clusters that had surface area,
removing total area, but these contributions average to zero.)
This argument implies the scaling relation
\be
\frac{\alpha - 1}{\nu} = d_s^c - d - \beta/\nu.
\ee
Note that this relation is consistent with modified hyperscaling,
\eqref{hyper}, and
the conjectured relationships among the domain wall dimensions,
Eqs.\ (\ref{eqnds}) and (\ref{eqnds2}).
Applying this result to our data, we find
\be
(\alpha -1)/\nu = -0.74 \pm 0.02,
\ee
giving our best estimate
\be
\alpha = -0.01 \pm 0.09.
\ee
Note that the magnitude of the $\beta/\nu$ contribution is small
compared with the error.

Our result for $\alpha$ is in marked disagreement with the
value from HY.
The scaling
assumptions for our and HY's analysis are identical.
It may be that one set of results is more strongly affected by finite
size errors, though we do fit larger values of $L$.
We note that the value of $\alpha$ that we find using $E_J$ is
extremely sensitive to the assumed value of $h_c$ and that the
uncertainty in $h_c$ dominates the error estimate. A change of
$h_c$ by $\delta h_c = 0.01$ gives a change
$\delta[(\alpha-1)/\nu]\approx 0.2$
or $\delta \alpha \approx 0.3$.
We are fitting for values near $h_c$, whereas the peaks in $C$ found
by numerical differentiation are somewhat above $h_c$.
In \secref{gs}, it is found that the convergence to a scaling function
for $h-h_c$ more than a couple of times $L^{-1/\nu}$,
where the peaks in $C$ are, is slow compared
with the convergence at $h=h_c$.

We use here two
independent data sets to arrive at our estimates for $\alpha$:
(a) total stiffness
measurements on isotropic and anisotropic samples, with fixed BC's on
two walls, applying finite size scaling,
and (b) the measurements of the bond part of the total energy,
$E_J$, using periodic isotropic samples, and fitting using a finite
size scaling form. For (b), we analyze the samples in two ways:
directly extracting $E_J$ data and also estimating the asymptotic
scaling using the $d_s^c$ and $\beta/\nu$
measurements. This latter method is least
sensitive to uncertainties in $h_c$.

\subsection{Deviations from criticality}

As the system is taken away from the critical point,
the nature of the spin clusters and correlations changes in
a straightforward way.  
If the exchange is increased, driving the system into the ordered phase,
then the large inverted droplets, which typically have gained energy
of order $\ell^\theta$ at
the critical point, will usually have this energy gain overcome by the extra
exchange energy cost of order $\delta J \ell^d_J$ when
$\ell$ is greater than the correlation
length, $\xi$.  Large inverted regions will be exponentially rare on
length scales longer than $\xi$.  

If the random field is increased or the exchange decreased to drive
the system into the disordered phase, 
we can no longer simply focus on the inverted regions that
already exist at the critical point but must also consider putative 
inverted regions that {\it could} exist.  In any region with
diameter of order $\ell$, there will be, at the critical point,
an excitation that flips of order $\ell^d$ spins for a typical
energy cost of order $\ell^\theta$ (more
precisely it will only flip of order
$\ell^{d-\beta/\nu}$ because of the frozen regions within it
which are not sensitive to the boundary of the region). 
Since decreasing $J$ will decrease the energy cost of this excitation
by an amount of order $\delta J L^{d_J}$, a good fraction of these
``excitations"   will have
negative energy and thus occur spontaneously at scales of
order $\xi$.  On this and larger scales, the orientation of the spins
will be determined primarily by the
local random fields within a distance of order $\xi$ of the spins
of interest.

\subsection{Thermal fluctuations and excitations}

The effects of thermal fluctuations have been discussed elsewhere
\cite{FisherRFIM,Fisher-act-dyn-scaling} in the general framework of a
zero-temperature random field critical fixed point.  We will thus
restrict ourselves here to a few comments in light of the present more
detailed picture.

At the critical point, as has been outlined above, there should be
potential excitations with energy of order $\ell^\theta$ around each
point, an independent one for roughly each factor of two in length
scale, $\ell$.  Since the energies of these are random there is a
finite probability density that the energy of any given one of them is
near zero --- indeed there would have been ones with negative energy
but these give rise instead to the inverted clusters in the ground state.  The thermal
fluctuations are dominated by the {\it rare active excitations} whose
energy is within of order $T$ of zero.  Because $\theta$ is positive, the active excitations with diameters of
 order $\ell\gg 1$  occupy only a small fraction -- of order
$T/\ell^\theta$ -- of the volume. But this small active fraction
dominates the correlations, in particular causing the thermal
fluctuations of the spin-spin correlations, the {\it truncated
correlations}, to decay as
\be
\overline{\langle (s_x-\langle s_x \rangle)(s_y-\langle s_y\rangle)\rangle} \sim \frac{T}{|x-y|^{d-2+\eta}}
\ee
where the exponent $\eta$ is related to that of the
zero-temperature correlations \eqref{zero-correlations} by
\be
\eta = \tilde{\eta} + \theta
\ee
the extra factor of  $T/|x-y|^{\theta}$ coming from
the probability that both spins are in the same
active excitation.\cite{FisherRFIM}
 
In general, except for fluctuation quantities such as the truncated correlations, the statements
that we have made about zero temperature will hold with minor (if sometimes subtle) modifications 
provided one considers always  free energies instead of energies.

One effect which must be mentioned, however, is the
``hypersensitivity" to changes along the critical line --- 
sometimes, rather misleadingly, referred to as ``chaos".
As long as $\theta < d/2$, which we believe
it probably is, although only barely so, 
which spins have which
orientation at the critical point will depend,
on sufficiently large scales, extremely sensitively on 
where one is on the critical line.\cite{BrayMoorestates,AlavaRieger}
Unfortunately, due to the smallness of $d/2 -\theta$,
this effect is unlikely to be observable in three dimensions 
but may be in higher dimensions for which $\theta$ is
expected to deviate more significantly from $d/2$. (In six dimensions and above, $\theta=2$.)

\section{Ground states and sensitivity to boundary conditions}\label{gs}

The simple picture of the random field Ising system exhibits two
phases with a single transition between them: an ordered phase in
which a typical spin is aligned with others far away; and a disordered
phase in which the magnetization is zero and the orientation of each
spin is determined locally by the random fields in its vicinity.  In
the ordered phase, $h<h_c$, spins have long range correlations and
there are both ``up" and ``down" states, although domain walls can be
introduced that divide the system into up and down regions.  In
contrast, when $h > h_c$, the spin correlation function is short
ranged, with characteristic scale $\xi \sim (h-h_c)^{-\nu}$ and there
is only one state; because of the locality, large scale domain walls
do not exist in this phase.

But it is interesting to ask, by analogy with spin glasses and other
systems with quenched randomness, whether the random field Ising
system could be more complicated, especially near to the critical
point. In order to address this, we must characterize the
macroscopically distinct states in an infinite system: is there, as
the simple picture would suggest, simply one state in the disordered
phase and two in the ordered phase?  Or is the behavior more subtle?

It has been claimed in the literature that ``replica symmetry breaking''
calculations show the existence of
an intermediate glassy phase, where many solutions with distinct
local magnetizations coexist
for a finite range of parameter, between the paramagnetic
and ferromagnetic phases.\cite{MezardYoung,MezardMonasson} But what
does this mean? Indeed, what does one mean by ``ground states" in an
infinite system with random couplings?  Furthermore, if one answers
these questions, what is the connection between multiplicity of
infinite system ground states and the notion of ``replica symmetry
breaking"?  

To consider these questions, it is simplest to restrict consideration
to systems, such as the RFIM with Gaussian random fields, in which the
finite system ground states for given boundary conditions are
non-degenerate with probability one. (Otherwise one gets into the
complications of ground state entropy as in diluted antiferromagnets
in a field and the bimodal
RFIM;\cite{BasteaDuxbury,HartmannRFIMdegen} but these issues
are distinct from the basic questions of ``states'' on which we focus.)

\subsection{Infinite system ground states} 

A {\it ground state of an infinite system} with finite range interactions
is a configuration whose energy cannot be decreased by changing {\it
any finite collection} of spins.  Equivalently, a ground state can be
thought of as the limit of a sequence of finite system ground states
of larger and larger subsystems, generally with appropriately chosen
boundary conditions on each size.  Thus the set of all ground states
for a specific infinite system, is the set of all distinct limits of
sequences of boundary conditions.\cite{NewmanStein} For two
ground states to be distinct, they must be distinguishable within some
finite distance of the origin: if the finite system ground states
differ only in regions whose distance from the origin grows without
bound as the system size increases, then the infinite system ground
states are the same.\cite{NewmanSteinComment}  All infinite system ground states have the {\it
same energy density} but comparing the energy of a pair of ground
states is not generally well-defined.

Many of the subtleties involved in considering infinite system ground
states come to the fore in the ordered phase of the random field Ising
model.  If we take the limit of larger and larger systems with open
(i.e. free) boundary conditions centered, for example, on the origin,
then the finite system ground states will not approach a limit! This
can be readily understood in terms of the ``up" and ``down" states
which we know exist in the infinite system -- albeit with some finite
density of misaligned spins.\cite{proofs} A given finite sample of
volume $\cal{V}$ will typically have random fields whose net effects
are to cause an energy difference between the up and the down states
which is of order $h\sqrt{\cal{V}}$.  Thus the ground states with open
boundary conditions will alternate randomly from mostly up to mostly down as a function of (the logarithm of the) system
size.  Of course, the up and down states can be found
by either taking the appropriate subsequences with open boundary
conditions, or by taking $+$ or $-$ boundary conditions on all sizes.
The problem of energy comparison is now clear: which of these two
states has the lower energy in a specific infinite system?  This is
manifestly ill-defined, indeed, because of the effects of the boundary
conditions, it is not possible to uniquely define the energy of an
infinite system ground state to higher accuracy than of order the
surface area of the region under consideration.

We can, however, compare the energies of {\it some} pairs of infinite
system ground states even in random systems.  In high dimensions,
greater than three, one can make ground states in the ordered
phase of the RFIM with a
domain wall that passes near the origin with a chosen orientation by
putting $+$ boundary conditions on half of the boundary and $-$ on the
other half.  If the random fields are weak enough (in four and five
dimensions, or with arbitrary randomness in the ordered phase in six
or more dimensions), the domain wall will be flat on large scales with
only {\it finite} typical deviations from planarity and its position and
orientation can then be fixed by its intersection with the boundary
which is forced  by  a ``seam" between  $+$ and $-$ areas of
the boundary conditions. The infinite system domain wall state so
constructed will be stable to changing any finite collection of spins,
but, in a well-defined sense, it has higher energy than either the up
or the down states. As the domain wall costs energy per unit area, if
one looks at a sufficiently large region that overlaps the domain wall
--- say cubical with $v=\ell^d$ --- the difference in energy
between the domain wall state and the up state will be of order $J
\ell^{d-1} \pm h \ell^{d/2}$ which is positive almost surely in the limit of
large $\ell$.

In contrast to the higher dimensional case, in the three dimensional
RFIM of primary interest, one {\it cannot} make infinite system domain
wall states straightforwardly even in the ordered phase.  If one tries
to set up a domain wall that is, say, horizontal in a system of size
$L\times L\times L$, one will find that the wall wanders in
the vertical direction away from the plane determined by the boundary
joint by a random, sample and subsystem size dependent amount of order $L^\zeta$ with
$\zeta=2/3$.\cite{GrinsteinMa,HuseHenley,FisherFRG} No matter how one
adjusts the boundary seam, one is unlikely, in the large system
limit, to be able to force the wall to be both near the origin and nearly
horizontal. Thus the sequence of domain wall forcing boundary
conditions will, in the ordered phase, contain one subsequence which
converges to the up state, another which converges to the down state,
and, almost surely, no other convergent subsequences.  (There are
subtleties, which we will not go into here, if one allows a wall in the
ordered phase to have {\it any} configuration dependent
orientation; these will be addressed in
Ref.\ \onlinecite{tobepubAAM}.)

The crucial question that we would like to address here is whether
there exists more than one infinite system ground state either at the
critical point or  slightly into the disordered phase.  In principle, to
investigate this one would need to study all possible sequences of
boundary conditions, obviously an impractical task.  In practice, one
must restrict consideration to some small subset of boundary
conditions and try to extract useful information about the infinite
system limit by carefully studying the size dependence of various
boundary conditions on regions near the origin.

\subsection{Numerical studies}

We have studied how the ground state configurations change in response
both to varying the boundary condition at fixed size and
to changing the system size.
We compare configurations for which the boundary spins are ``open'' ($O$),
fixed positive ($+$), fixed negative ($-$), and random fixed spins ($R$).
For fixed size calculations, for each realization of the random fields we compare all possible pairs of
boundary conditions in the set $\{O,R,+,-\}$.
We also compare ground state configurations for
open boundary conditions on
a sample of size $2L-1$ (denoted $D$) that contains a subsample
of size $L$, with the states for 
boundary conditions $O$, $+$ or $-$ imposed on the subsample.
(The values of $L$ were taken to be odd for these comparisons,
so that the origin coincides with a spin.)
The results of all of the comparisons are characterized by
counting how many spins differ for the two
boundary conditions in a volume $w^3$ centered at the origin.

The primary emphasis of these calculations is to determine whether changes in boundary
conditions can create configurations that differ from those with uniform
$+$ or uniform $-$ boundary conditions, i.e., those that produce the up and the down states in the ordered phase.
If the $+$ and the $-$ boundary conditions produce identical
configurations in the deep interior, this suggests that there is only one state. 
If the probability that some other boundary condition produces a configuration
in the interior  that differs from those of  {\it both} the the $+$ and $-$
boundary conditions, vanishes as $L\rightarrow\infty$, this suggests
that there are at most two states.

We report here a selection of results
for
(a) the probabilities $P_{O\pm}(h,w,L)$ ($P_{R\pm}(h,w,L)$) that
the boundary condition $O$ (respectively, $R$)
gives a central volume $w^3$ that
differs from that for {\em both} $+$ and $-$ boundary conditions at
fixed sample size $L$;
(b) the probability $P_{DO}(h,w,L)$ that
the number of differing spins within the window is non-zero when one
compares open boundary conditions for samples of size $2L-1$
and a subsample of size $L$;
and
(c) the probability $P_{D\pm}(h,w,L)$ that
open boundary conditions on the larger sample
gives a central volume $w^3$ that
differs from that for {\em both} $+$ and $-$ boundary conditions on
the smaller sample.

\begin{figure}
\centering\includegraphics[width=8.cm]
{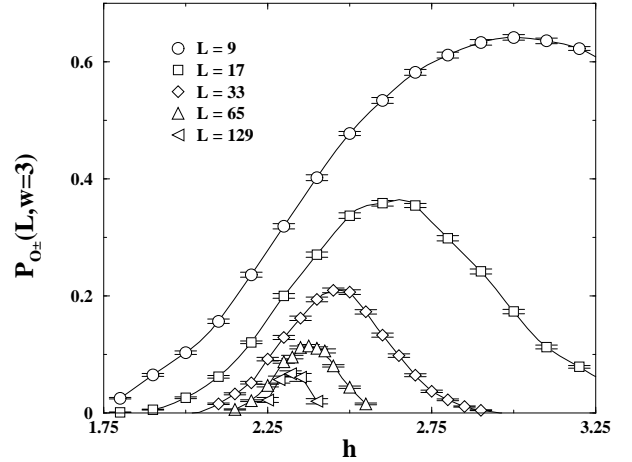}
\caption{
Plot of the unscaled probability $P_{O\pm}$
that the central window
of size $w=3$ of a ground
state configuration with open BCs on a given sample of size $L$, differs 
from the  configuration in the window with
{\em both} uniform $+$ and $-$ fixed boundary
conditions.
The lines are intended
to organize the data visually.
}
\label{fig_of_noscale}
\end{figure}

\begin{figure}
\centering\includegraphics[width=8.cm]
{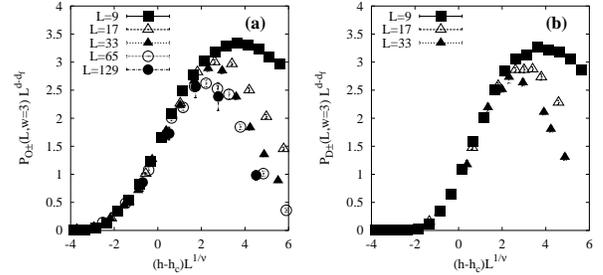}
\caption{
Scaling plot for the probability that the central window
of size $w=3$ of a ground
state configuration 
differs from that of uniform $+$ or $-$ fixed boundary
conditions for (a) open boundary conditions on the same sample of
size $L$ and
(b) open boundary conditions on a sample of size $2L-1$.
The values used for scaling are $h_c = 2.270$, $\nu = 1.37$, and
$d_f = 2.25$. The probabilities scale very well near $h=h_c$, but the
peak heights, at $h>h_c$,  converge slowly.
}
\label{fig_ofrf}
\end{figure}

The calculation of the probabilities $P_{O\pm}$ and $P_{R\pm}$
(comparisons (a)) allows us to study ground states near $h_c$.  The
events of interest are those where a given boundary condition $B$,
either $B=R$ or $B=O$, gives a configuration distinct from both the
$+$ and the $-$ boundary conditions. For $h>h_c$ as
$L\rightarrow\infty$, $P_{B\pm}$ is expected to go to zero, since the
effects of the boundary penetrate only a distance $O(\xi)$ into the
sample. For $h<h_c$, in contrast, most of the interior configuration
is either $+$ or $-$ and the chances that random or open boundaries
yield some other possibility should again decay exponentially.  At
$h=h_c$, the correlation length diverges and at this critical point,
we expect that the probability of a domain wall passing near the
center decays only as a power of the system size.  Using simple
arguments based on the fractal nature of domain walls,
\cite{AAMstates,PalassiniYoung2d} the probability that a window of
size $w$ will intersect an object of fractal dimension $d_f$ scales as
$(w/L)^{d-d_f}$. This is analogous to the probability of a domain wall
in the ordered phase passing near the origin as discussed above. The
appropriate fractal dimension to use here at the critical point is the
dimension from box-counting, $d_I$, that we studied above. Basically,
there is a substantial probability that open or random boundary
conditions will, at the critical point, induce a system spanning
domain wall relative to the $+$ and $-$ boundary conditions.  Near the
critical point, scaling suggests the form
\be
P_{B\pm}(h,w,L) = L^{d_f - d} {\cal P}_B[w,(h-h_c)L^{-1/\nu}]
\ee
for $B=R$ or $O$. We plot our data for $P_{O\pm}$, $w=3$, in 
\figref{fig_of_noscale} and \figref{fig_ofrf}(a), assuming
this scaling form, taking
$h_c = 2.270$, $d_f=2.25$ and the best fit value $\nu=1.37$.
The results for $P_{R\pm}$, while not shown here, are nearly identical,
apparently converging for large $L$  near the critical point to an
extremely similar, if not the same, scaling
function, though the smaller $L$ curves have slightly different finite
size corrections.
We expect that $d_f$ is equal to the incongruent domain wall
dimension $d_I$, as this is the domain wall dimension that describes
changes in the bonds, and this
expectation is consistent with our results.

We note that taking the value $d_f=2.20$  appears to give a better  fit for
the peak heights away from $h_c$, but as convergence in several quantities
is poorer away from $h_c$, the  value $d_f=2.25$ is acceptable.
Directly fitting {\em the peaks} for $d_f$ gives a value $d_f=2.22\pm 0.03$.
 
The data for comparisons (a) should also scale with $w$ for large $w$:
${\cal P}[w,(h-h_c)L^{-1/\nu}] = w^{d - d_f} {\cal P}'[(h-h_c)L^{-1/\nu}]$.
However, we do not have enough range in $w$ for $w>>1$ to confirm this;
for $w$ small, discreteness effects will prevent a collapse.
For $L=129$, the data do collapse 
well for $w=65,33,15$, assuming the above
scaling form and the best fit values of $h_c$, $d_I$, and $\nu$.
Note that similar finite $w$
effects were also seen in the data of Ref.~\onlinecite{AAMstates}, where
large $w$ was needed to see convergence to power law behavior in $w$,
though scaling worked well for fixed $w$ with $L \gg 1$.

\begin{figure}
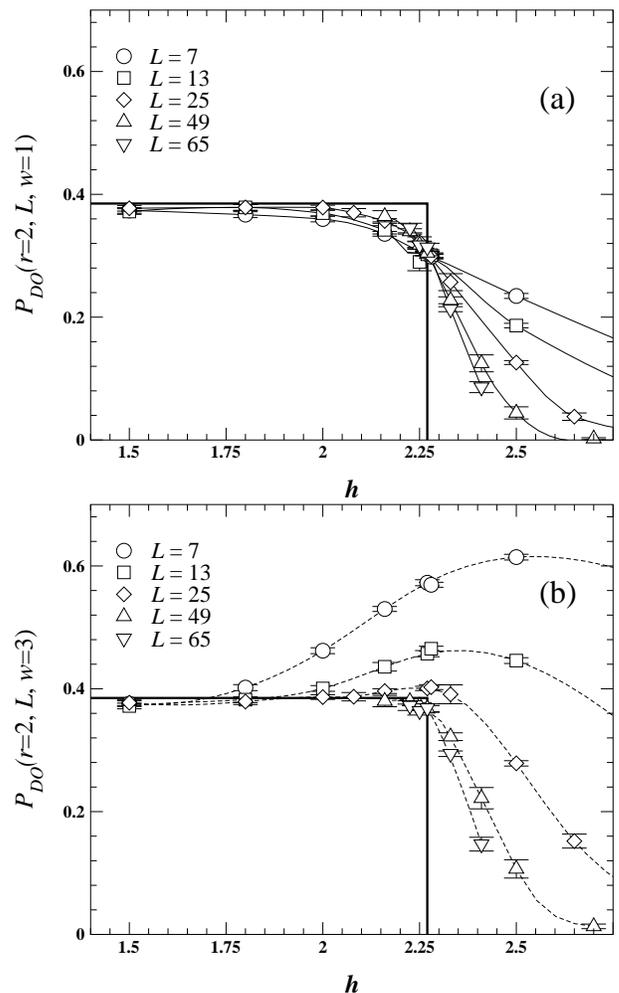

\centering\includegraphics[width=8.cm]
{30a.eps}
\centering\includegraphics[width=8.cm]
{30b.eps}
\caption{
Probability $P_{DO}$ that the central window
of size (a) $w=1$ and (b) $w=3$ of a ground
state configuration in a subsample of size $L$
differs from that in a sample of size $2L-1$, with open
boundary conditions on the subsample and sample.
Note that the sample sizes are approximately separated by a factor of $2$,
except for the largest two sizes.
For small $h$, the probability $P_{DO}\approx 0.38$, quantitatively consistent
with a simple model with two states.
For large $h$, $P_{DO} \rightarrow 0$
as $L\rightarrow\infty$, consistent with a single state.
The solid line shows the step function that would obtain
for $P_{DO}$ in the $\infty$-volume (and large $w$)
limit, {\em if} it were the case that $|m|\equiv 1$ in the
ferromagnetic phase, taking $h_c = 2.270$.
The data is consistent with the calculated $P_{DO}$ values
approaching this step function at larger sizes $L$, for $h \neq h_c$.
Note that at $h=h_c$,
$P_{DO} \approx 0.368 \pm 0.006<0.379\ldots$ for
$L=97$ and $L=129$, $w=3$, consistent with a constant or slowly
decaying value of $P_{DO}$ at the critical point.
The dashed curves are spline fits 
to organize the data visually.
}
\label{fig_P2}
\end{figure}

Comparison (b) compares open
boundary conditions on two samples of
different sizes, the smaller being a
subsample of the larger centered at the origin.
In the disordered phase, with the local spin
configuration determined by the random
fields nearby,  doubling the size of the system is not expected to
change the configuration in small windows near the origin,
for $w \ll L$ and $L \gg \xi \sim (h-h_c)^{-\nu}$. But for $h<h_c$ and
$L\gg\xi\sim(h_c-h)^{-\nu}$, the
spin orientation is determined by the sign of the
total (effective) random field which
will depend stochastically on the system size as discussed in the previous subsection.
For a fully  magnetized system, ($|m| = 1$),
these simple expectations yield $P_{DO} \rightarrow 0$
for $L\rightarrow \infty$ with $w$ fixed for $h > h_c$, and
\begin{equation}
P_{DO} \rightarrow (7\pi^2)^{-1/2}\int_{0}^\infty dx\,
\int_{0}^{\infty} dy\, e^{(-y^2/2 - (y+x)^2/14)} \label{doubleint}
\end{equation}
for $h < h_c$ as $L\rightarrow\infty$.
The integral in Eq.\ (\ref{doubleint})
is the probability that the total random field in
the volume $(2L)^3$ exceeds in magnitude and is opposite in sign to
the total random field in a subvolume $L^3$,
assuming Gaussian distribution of the
field on length scale $L$ with variance $L^3$.
This integral gives a value $P_{DO}= 0.384973\ldots$ for $h<h_c$.
The results of our ground state studies, displayed in \figref{fig_P2}, appear to be 
consistent with this  limit, for $h<h_c$. This confirms the expectation that as the infinite volume
limit is taken in the ordered phase,
the spins in a fixed volume flip between two distinct configurations
infinitely often --- typically every factor of three or so in length scale.
Near the critical point, the probability of differences in
the window between the full 
and the subsample will be modified since with $|m|< 1$, there is a non-zero
probability that the window will be contained in a
frozen spin clusters that is 
unaffected by the overall majority random field.
But this $w$-dependent difference
only becomes important near $h_c$, as $\beta$ is so small. 

Right at the critical point the effects of frozen clusters
on all scales should in principle suppress
$P_{DO}$  to zero in the large
$L$ limit for all
$w$; but as it will decay only as $1/L^{\beta/\nu}$, this effect
is hard to see. 
In the disordered phase, our data is consistent with
$P_{DO}$ vanishing exponentially for $L\gg \xi$. 

Comparison (c) allows us to address nearly the same question as (a),
but more directly checks that increasing the volume of the system has
the effect of setting an effective boundary condition of $+$ or $-$ on
the central region. The scaling collapse, shown in \figref{fig_ofrf}(b)
is acceptable, as it is in (a), with a scaling function similar to, but
distinct from that for comparison (a). The results for $P_{D\pm}$ show
that, except for a region near $h_c$ that shrinks and decreases in
probability with increasing $L$, the configuration given by the larger
system with open boundary conditions does {\it not} produce a distinct
interior volume from that found by imposing $+$ or $-$ boundary
conditions on the smaller system of size $L$.

Taken together, these results are consistent with the expectations
from the simple scenario for the structure of the states given above;
there do not seem to be any indications of stranger behavior.  Thus,
in the absence of concrete testable predictions from those who believe
there should be more than just the simple set of states, we can do no more
than conclude that if they can indeed occur, it must be only under very
subtle conditions.

\section{Summary}\label{summ}

In this paper, we have presented numerical results for the ground
states of 3D random field Ising magnets focusing on the transition
between the ordered and disordered phases.  Our results allow us to
conclude that the transition is second order, though the magnitude of
the magnetization vanishes very slowly as the critical random field
strength, $h_c$, is approached from below. In addition to the
magnetization, we have studied the stiffness of the system and some of
the geometrical aspects, in particular the fractal properties of
domain walls at the critical point.  In general, the results agree
very well with a scaling picture of the transition introduced some
time ago \cite{FisherRFIM} and extensions of it to the properties
studied here.

Some earlier authors have suggested that the behavior of the RFIM near
to the ordering transition will be more complicated than this
scenario, arguing for some kind of ``replica symmetry breaking".
Although, as is so often the case, the meaning of this term in this
context has not been made clear, if we take it to imply the existence
of many infinite volume ground states, this would have testable
consequences.  Although a full test of the dependence of the ground
states on sequences of boundary conditions that this would imply is
beyond the scope of today's computers and algorithms, we have made
some preliminary tests on the dependence on boundary conditions.  In
particular, we have studied the probability that the configuration in
a fixed volume at the center of a sample can be induced to differ from
both the fixed $s=+1$ and fixed $s=-1$ boundary conditions by various
other boundary conditions.  With the range of boundary conditions we
have tested, this probability vanishes in the expected manner as
$L\rightarrow\infty$. Indeed, the power law dependence of this
probability on $L$ and the scaling with $h-h_c$ are consistent with
the domain wall fractal dimension and correlation length exponents
determined by other methods. Our results are thus consistent with a
single disordered to ordered transition at $h_c$, with a unique state
in the disordered phases and a pair of states (``up" and ``down") in
the ordered phase.
Recent simulations at finite temperature in smaller
systems by Sinova and Canright,\cite{SinovaCanright}
who used the spectrum of the spin-spin
correlation matrix and $P(q)$ distributions, also suggest a single
transition.

At non-zero temperature, the thermodynamic properties of the phase
transition are believed to be similar to those at zero temperature:
the transition is governed by a {\it zero temperature fixed point}.  But at
positive temperature, one can also consider dynamic effects; indeed,
as has been known for a long time, these dominate both Monte Carlo
simulations and experiments.
As first pointed out by Griffiths,\cite{Griffiths} random
systems can have singularities --- albeit very weak ones --- in
thermodynamic properties well before the transition is reached and
this will be the case for the RFIM.  These rare region effects are
unobservable as far as equilibrium properties in classical systems, but
do have dynamic
consequences.\cite{Randeria-Sethna-Palmer,Dhar,Huse-Fisher-rare}
In the Griffiths region above the
transition, the average dynamic autocorrelations will decay more
slowly than exponentially because of the effects of anomalously
ordered local regions.  Perhaps this kind of rare-region effect, and
the more interesting but related effects that occur as the transition
is approached, are all that is meant by ``replica symmetry breaking".
If this is the case, then it would be nice if the proponents of these
ideas would say so.  If not, then it is incumbent upon them to come up
with some testable predictions.  If these can be tested by static
ground state properties, the RFIM is as good as
system as any on which to perform such tests as
the system sizes that can be studied are quite impressive: comparable
to the largest that can be studied by Monte Carlo simulations in pure
systems.

\begin{acknowledgments}
AAM would like to thank Jon Machta for stimulating
discussions. AAM and DSF would like to thank Alexandar Hartmann and
Peter Young for discussions of their recent results.
This work was supported in part by
the National Science
Foundation via grants DMR-9702242, DMR-9809363, DMR-997621, and
DMR-0109164, as well as by
the Alfred P. Sloan Foundation, and a grant of computing time from NPACI.
\end{acknowledgments}

\begin{appendix}\label{appA}
\section{Algorithm implementation}

We briefly describe here the algorithm and code  used, including
modifications to the conventional RFIM max-flow problem; outline the
verification procedure for the code; and briefly outline the statistics
and error bar procedure.

\subsection{Base code and modifications}

There is a now well-known mapping of the RFIM ground state problem
to a min-cut/max-flow problem. \cite{dAuriac} This correspondence and
the push-relabel algorithm for the max-flow problem,
including terms used here (such as layers and excesses),
is well described in reviews and
texts, such as Refs.\ \onlinecite{RiegerReview}, \onlinecite{ADMR},
and \onlinecite{combopt}.
The implementation of the Goldberg-Tarjan \cite{GoldbergTarjan}
max-flow algorithm that we started with was the \texttt{h\_prf}
code in C written by Cherkassky and Goldberg, \cite{Goldberg} which
in general performs quite well for a number of graph topologies.

We modified the code to be more compatible with the C++ language
and developed objects (including samples, configuration subsets as windows,
and random number generators) to conveniently implement
a variety of boundary conditions and analyses. One very simple benefit
of an integrated code is that the graph input, which is quite costly
when read as a text file, is greatly sped up. More importantly, the
short main routine was easily modified to compute answers to a wide number
of questions.

The most significant change to the core push-relabel code was a
modification that allowed for positive and negative excesses.
This modification was developed in collaboration with D.\ McNamara.
\cite{mcnamarathesis} The central idea is the elimination of the
source and sink nodes, which conventionally have links to the nodes
of the graph representing the spins $s_i$, in favor of
introducing nodes with a negative excess. The first step in the
conventional algorithm pushes as much flow as possible from the
source onto the lattice nodes. This step is replaced in our code
with an initialization where a positive
excess $h_i$ is placed on each node for which $h_i > 0$.
The connections to the sink are substituted for by placing a {\em negative}
excess on the nodes with $h_i < 0$. The push-relabel algorithm then
proceeds with the usual steps, with the nodes with positive excess
having their excess pushed and their heights relabeled, as appropriate.
The negative excess nodes act as sinks for the positive excess, until
such a nodes total
excess becomes positive. Besides removing the links to the source and
sink, the global relabeling step must be modified. Instead of carrying out a
breadth first search from the sink node, the breadth first search instead
starts
from the nodes with negative excess. (If no negative excess nodes remain,
the algorithm terminates with flow equal to the sum of the positive $h_i$.)
The initial totals of the positive and the negative excesses are compared
with the final totals: the decrease in the total positive flow, for example,
gives the maximum flow through the graph. The spin configuration
and magnetization is determined
by counting the number of nodes that are in the maximal layer.

The removal of the source and sink nodes reduces the amount of memory
used by an amount $1/(d+1)$ relative to the conventional memory
requirements and results in a slight speedup. For the largest lattice
sizes studied ($256^3$), memory requirements were reduced at the cost
of speed. If pointers and integers each require 4 bytes,
the Cherkassky and Goldberg implementation requires 16 bytes for each arc
and 32 bytes for each node (counting the layers as part of the per-node
requirements.) The use of pointers was retained for system sizes
up to $128^3$. For a regular lattice, however, the nodes
at the end of each arc, sister arcs, and the list of arcs at each node
can be recomputed whenever needed. For a cubic lattice, then, the
number of bytes per node is reduced from $(6\times 16 + 32)=128$ bytes
to $(6\times 4 + 32) = 56$ bytes. For $L^3=128^3$ samples, the running
time increased by a factor of $\approx 2.5$, primarily
due to the recomputation of the tail nodes of the arcs and the sister
arcs.

One modification for the $256^3$ samples
was made that is not strictly sound, in that the
algorithm could conceivably fail. In order to save memory, a limit
to the maximum number of layers was implemented. In the
Cherkassky and Goldberg code, the number of layers allocated is given by the
number of nodes in the graph. In practice, however, far fewer are
needed.
A check over $1000$ samples for $8 < L < 128$ was carried out, for
several values of $h$.
The number of layers needed, $K$, appears to be largest for $h\approx h_c$.
At this value, the sample mean of the maximum layer needed is about
$\overline{K} \approx 2L$.
The distribution over samples of the number of layers needed
roughly scales with $L$, though the dependence of the
width of the distribution could be $L^x$, with $x$
near 1. In any case, the distribution drops off very quickly with $K$.
The maximum number of layers needed over $1000$ samples scales roughly
linearly with $L$, $K^{\rm max}\approx 7L$, where the maximum is for
periodic boundary conditions and over a range of $h$, $2.0<h<2.5$,
with a peak in $K^{\rm max}(h)$ near $h_c\approx 2.27$ (though the 
peak is slightly above $h_c$ for smaller samples.)
The mean number of layers fits relatively well to a scaling collapse,
with a maximum value scaling consistent with $L^{1.08}$ (or even $L\sqrt{\ln L}$),
scaling about $h_c\approx 2.27$
with $\nu\approx 1.35$.
We set $5 \times 10^4$ as the maximum number of layers for all
sample sizes, which is
nearly $200\times L$ for the largest samples studied.
This number of layers was easily sufficient for all samples studied.
The amount of memory needed
for cubic lattices is then $48 + O(L^{-2})$
bytes per node.

\subsection{Verification}
The modifications made to the base code, while theoretically
sound (except for the limit on the number of layers),
could inadvertently introduce errors, due to errors in coding. We therefore
verified the code against the Cherkassky and Goldberg
codes \texttt{h\_prf} and
\texttt{hi\_pr} (version 3.3) codes
\cite{Goldberg,starlabs} and a selection of other codes that were
not based on a push-relabel algorithm. This was done by having the
production code write out the list of the $h_i$. A small program
then converted these $h_i$ into arcs and nodes into a graph description
DIMACS format, using the conventional
representation with a source and a sink.

These graph descriptions were then used as input
to \texttt{h\_prf}, \texttt{hi\_pr}, and other codes, such as those
developed in the First DIMACS Challenge.
\cite{dimacschallenge1} The flows and the magnetization from these
available algorithms were then
compared, sample by sample, with the production code. The precise
comparison was done for a few tens of samples with ferromagnetic coupling
strength $J=10,10^2,\ldots,10^7$, relative disorder strength
$h/J=1,2.3,3,10$ and system sizes $L=4,8,...,128$. The production code
and the other algorithms agreed in all cases, except when the flow exceeded
$2^{31}$, which was generally the maximum possible flow in the available
algorithms. The production code used here does not have that restriction,
as flow computation is done at the end by comparing the initial and final
positive and negative excesses, which were summed up as double precision
quantities. The code was able to handle larger flows consistently, as could
be verified by scaling $h$ and $J$ to large values (multiplying $J$
and the $h_i$ by a factor of,
say, $10^5$, and checking that the maximum flow increases proportionately and
that the configuration is unchanged.)

For efficiency, we have used an integer algorithm, with a resolution
of $10^4$ by replacing the $h_i$ by random integers found by rounding
$z_i \times 10^4$ towards zero, with $z_i$ Gaussian random variables
with zero mean and unit variance and the exchange by $J/h \times
10^4$. We checked that {\em reducing} this resolution by a factor of $10$
for selected measurements did not affect the computed averages, such
as the stiffness in the isotropic and anisotropic
samples for systems up to $128^3$. For a resolution of
$10^2$, there were discrepancies outside of statistical errors, but
these discrepancies could be consistently explained by the effects of
rounding to an integer, which shifts the width of the distribution of
$h_i$ by a small, predictable amount.

We also tested the assumption that the cut found was unique (that is,
that the ground state was unique, for a given sample.)
Some accidental degeneracies were found, at the level of a
fraction of spins $\approx 0.9 \times 10^{-6}$,
for $h$ near $2$, including $h_c$.
This would result in the magnetizations  being in error at the level of $<10^{-6}$,
well within the statistical errors.
Increasing the resolution by a factor of $10$ increases the running time
by about $7\%$ and reduces the fraction of degenerate
spins to $\approx 2\times 10^{-7}$.
As the degeneracies were for the most part attributable to single spins,
were rare, and did not affect any of the sample averaged results in the
cases we tested, the integer
scale of $10^4$ was more than sufficient for this study.

We have verified that our choice of
random number generator does not affect the results. Specifically, we
used two generators for the computations of the magnetization and
domains (defined in \secref{clust}) in $256^3$ samples at $h = 2.27 \approx
h_c$ and found the results to agree within statistical errors (the
results reported pool the results from these generators.) We also
checked the results from the two generators against each other for a
larger number of smaller systems.

Though quantities computed and the
details of our interpretation differ
from previous work, the numerical values for the same sample
sizes and measurements (for example, magnetization
and the largest cluster sizes)
are consistent with published data.\cite{HartmannNowak,EsserNowakUsadel}

\begin{figure}
\centering\includegraphics[width=8.cm]
{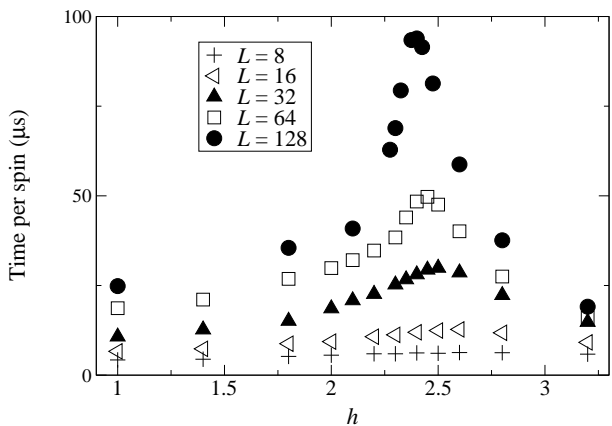}
\caption{
Elapsed time for computing ground states in the RFIM, plotted vs.\ $h$,
for linear sizes $L=8,16,\ldots,128$. The ``fast'' algorithm is
applied, with the larger memory requirements, on a 766 MHz Pentium III
processor. The peak times {\em per spin} scales nearly linearly with
$L$.
}
\label{fig_timing}
\end{figure}

\begin{figure}
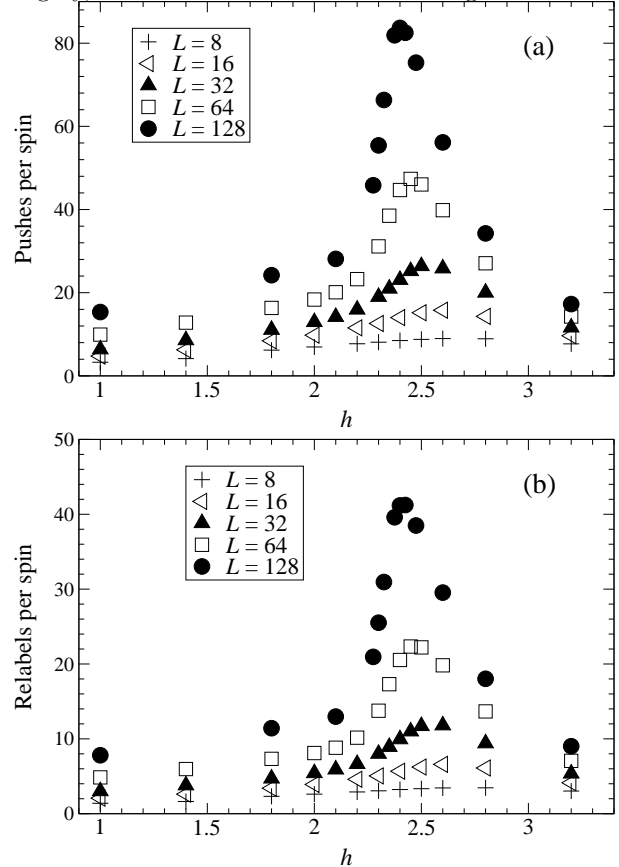

\centering\includegraphics[width=8.cm]
{32a.eps}
\centering\includegraphics[width=8.cm]
{32b.eps}
\caption{
Statistics for the ``operations'', pushes (a) and relabels (b),
performed in
computing ground states in the RFIM, plotted vs.\ $h$,
for linear sizes $L=8,16,\ldots,128$.
The peak number of operations {\em per spin}
scales nearly linearly with $L$ (near $h_c$). At both high and
low $h$, the number of these rearrangements scales
nearly as $L^{0.3}$.
}
\label{fig_pushrelabel}
\end{figure}

\subsection{Timing}

Consistent with similar optimization problems related to physical
problems, the typical CPU time needed to find the ground state
scales roughly as $N^{1.3}$ near $h_c$.
Roughly, it takes about 16-20 times longer
to find the ground state each time the sample size is doubled, for
$h\approx h_c$.
Using \texttt{CC} on a 400 MHz Sun UltraSparc II
(the San Diego Supercomputing Center Sun HDSC10000),
a $256^3$ lattice
required 913 MB of memory total for the graph data, the instructions,
and the data structures required
for analysis. Running time for this size and this architecture averaged
1.8 hours per sample, for $h=2.27$.
Run times, normalized to the elapsed time per spin, for the larger memory
algorithm, with the full data structure, are plotted in
\figref{fig_timing}. Clearly, the shape of the elapsed
time vs.\ $h$ sharpens some
as $L$ increases. The peak running time scales as $\sim L^{4.0}$
over the scales
$L=8$ to $L=128$.
Further details of the scaling of the running time and connections
between the algorithm and the physical concepts of ground state
degeneracy and correlation length are described in \refref{AAMphasepoly}.

\end{appendix}


\begin{thebibliography}{xx}

\bibitem{proofs}
J. Z. Imbrie, \PRL{53}, 1747 (1984); Comm. Math. Phys. {\bf 98}, 145 (1985).
J. Bricmont and A. Kupiainen, \PRL{59}, 1829 (1987).

\bibitem{FisherRFIM}
D. S. Fisher, \PRL{56}, 416 (1986).

\bibitem{Villain}
J. Villain, \prl{52}, 1543 (1984).

\bibitem{Ogielski}
A. T. Ogielski, \PRL{57}, 1251 (1986).

\bibitem{dAuriacSourlas}
J. C. Angl\`es d'Auriac and N.~Sourlas,
Europhys. Lett. {\bf 39}, 473 (1997).

\bibitem{Sourlas}
N.~Sourlas, Comp. Phys. Comm. {\bf 121}, 184 (1999).

\bibitem{MezardYoung}
M. M\'ezard and A.P. Young, Europhys. Lett. {\bf 18}, 653 (1992).

\bibitem{MezardMonasson}
M. M\'ezard and R. Monasson, \PRB{50}, 7199 (1994).

\bibitem{NattermannInYoung}
T. Nattermann in ``Spin Glasses and Random Fields'',
ed. A. P. Young (World Scientific, Singapore, 1998).

\bibitem{Swiftetal}
M. R. Swift, A. J. Bray, A. Maritan, M. Cieplak,
and J. R. Banavar,
Europhys. Lett. {\bf 38}, 273 (1997).

\bibitem{HartmannNowak}
A. K. Hartmann and U. Nowak,
Eur. Phys. J. B {\bf 7}, 105 (1999).

\bibitem{RiegerRFIM}
H. Rieger, \PRB{52}, 6659 (1995).

\bibitem{MachtaNewmanChayes}
J. Machta, M. E. J. Newman, and L. B. Chayes,
\PRE{62}, 8782 (2000).

\bibitem{dAuriac}
J. C. A. d'Auriac, M. Preissmann, and R. Rammal,
J. Phys. Lett. {\bf 46}, L173 (1985).

\bibitem{Goldberg}
B. Cherkassky and A. Goldberg,
Algorithmica {\bf 19}, 390 (1997).

\bibitem{combopt}
{\em Introduction To Algorithms},
T. H. Cormen, C. E. Leiserson, and R. L. Rivest
(MIT Press,
Cambridge, Massachusetts, 1990).

\bibitem{AAMrough}
A. A. Middleton,
\PRE{52}, 3337 (1995).

\bibitem{RiegerReview}
See, e.g., H. Rieger,
Lecture Notes in Physics 501 (Springer-Verlag, Heidelberg, 1998).

\bibitem{McMillanDWRG}
W. L. McMillan, \PRB{30}, 476 (1984).

\bibitem{BrayMooreLCRG}
A. J. Bray and M. A. Moore, \PRB{31}, 631 (1985).

\bibitem{BrayMoorestates}
A. J. Bray and M. A. Moore, \PRL{58}, 57 (1987).

\bibitem{AAMstates}
A.~A.~Middleton, \PRL{83}, 1672 (1999).

\bibitem{PalassiniYoung2d}
M. Palassini and A. P. Young,
\PRB{60}, 9919 (1999).

\bibitem{PalassiniYoung3dPRL}
M. Palassini and A. P. Young,
\PRL{83}, 5126 (1999).

\bibitem{2DFMSG} For the similar case of the ferromagnetic to spin
glass transition in two dimensions, see the numerical work of
W. L. McMillan, \prb{29}, 4026 (1984).

\bibitem{HartmannYoung}
A. K. Hartmann and P. Young, cond-mat/0105310.

\bibitem{JaccarinoBiref}
D. P. Belanger, A. R. King, V. Jaccarino, and J. L. Cardy,
\PRB{28}, 2522 (1983).

\bibitem{fit_rho_note}
The data is well fit assuming a simple correction to scaling
with $\rho(v) = \rho_\infty + b v^{-y}$, with $\rho_\infty = 0.0019$, to
within $10^{-4}$, $b$ a fitted constant,
and $y$ on the order of $0.4$.
The principle uncertainty in $\rho_\infty$ comes from the uncertainty
in $h_c$: the value to which
$\rho(v)$ converges would vary if $h_c$ is slightly greater or smaller
than $2.27$, as is apparent in \figref{fig_clustercount_cf_h2}.

\bibitem{inverse-square-Ising}
D. J. Thouless, Phys. Rev. {\bf 187}, 732 (1969);
M. Aizenmann, J. Chayes, L. Chayes, C. Newman,
J. Stat. Phys. {\bf 50}, 1 (1988).

\bibitem{CaoMachta} 
M. S. Cao and J. Machta, \PRB{48}, 3177 (1993).

\bibitem{NewmanStein}
C. M. Newman and D. L. Stein, \PRB{46}, 973 (1992);
\PRL{72}, 2286 (1994); {\bf 76}, 515 (1996);
{\bf 76}, 4821 (1996);
\PRE{55}, 5194 (1997); {\bf 57}, 1356 (1998).

\bibitem{FisherHuseStates}
D. A. Huse and D. S. Fisher, J. Phys. A {\bf 20}, L997 (1987);
D. S. Fisher and D. A. Huse, J. Phys. A {\bf 20}, L1005 (1987).

\bibitem{SofferSchwartz} M. Schwartz and A. Soffer, \PRL{55}, 2499 (1985).

\bibitem{ImryMa}
Y. Imry and S. Ma, \PRL{35}, 1399 (1975).

\bibitem{BrezinDeDominicis}
E. Br\'ezin and C. De Dominicis,
arXiv:cond-mat/0007457.

\bibitem{ChayesChayesFisherSpencer}
J. T. Chayes, L. Chayes, D. S. Fisher, and T. Spencer,
\PRL{57}, 2999 (1986); Comm. Math. Phys. 120, 501 (1989).

\bibitem{GrinsteinMa}
G. Grinstein and S. K. Ma, \PRB{28}, 2588 (1983).

\bibitem{HuseHenley}
D. A. Huse and C. L. Henley,
\PRL{54}, 2708 (1985).

\bibitem{FisherFRG}
D. S. Fisher, \PRL{56}, 1964 (1986).

\bibitem{ChauveLeDoussalWiese}
P. Chauve, P. Le Doussal, K. Wiese, \prl{86}, 1785 (2001).

\bibitem{tobepubAAM} A. A. Middleton, unpublished.


\bibitem{BC-pure-mag}
J. L. Cardy, J. Phys. A {\bf 17}, L385 (1984).

\bibitem{ShaoTu} J. Shao and D. Tu,
{\em The Jackknife and Bootstrap} (Springer-Verlag, New York, 1995).

\bibitem{EsserNowakUsadel}
J. Esser, U. Nowak, and K. D. Usadel, \PRB{55}, 5866 (1997).

\bibitem{depthnote}
The level of nesting of domains at $h_c$
varies with system
size. For $L=32$, $95\%$ of the samples had minority spin
clusters, but these had no subclusters; in the
remaining $5\%$ the depth $k=2$: minority clusters had majority spin inclusions.
Out of the $64^3$ samples,
$51\%$ had $k=1$ and $49\%$ had $k=2$.
About $0.1\%$ of the $128^3$ samples had depth $k=3$, $1.7\%$ had $k=1$,
with the remainder having $k=2$.
The $256^3$ samples had the highest
nesting, though as $\beta/\nu$ is small, the nesting is
still relatively small: $10$ samples out of $1000$, or about $1\%$, had
$k=3$, $989$ had a depth of $2$ and $1$ had $k=1$.

\bibitem{conf-invariance}
J. L. Cardy, J. Phys. A {\bf 17}, L961 (1984).

\bibitem{RiegerYoung} H. Rieger and A. P. Young, J. Phys. A
{\bf 26}, 5279 (1993).

\bibitem{Fisher-act-dyn-scaling}
D. S. Fisher, J. App. Phy. {\bf 61}, 3672 (1987).

\bibitem{AlavaRieger}
M. Alava and H. Rieger,
Phys. Rev. E {\bf 58}, 4284 (1998).

\bibitem{BasteaDuxbury}
S. Bastea and P. M. Duxbury,
Phys. Rev. E {\bf 58}, 4261 (1998).

\bibitem{HartmannRFIMdegen}
A. Hartmann, Physica A {\bf 248}, 1 (1998).

\bibitem{NewmanSteinComment}
D. Newman and C. Stein, cond-mat/0103616.

\bibitem{SinovaCanright}
J. Sinova and G. Canright, cond-mat/0103071.

\bibitem{Griffiths}
R. B. Griffiths, \PRL{23}, 17 (1969).

\bibitem{Randeria-Sethna-Palmer}
M. Randeria, J. P. Sethna, and R. G. Palmer,
\PRL{54}, 1321 (1985).

\bibitem{Dhar}
D. Dhar, M. Randeria, and J. P. Sethna,
Europhys. Lett. {\bf 5}, 485 (1988).

\bibitem{Huse-Fisher-rare}
D. A. Huse and D. S. Fisher, \PRB{35}, 6841 (1987).

\bibitem{ADMR}
M.J. Alava, P.M. Duxbury, C. Moukarzel and H. Rieger, 
{\em Exact combinatorial algorithms: Ground states of disordered systems},
For Domb and Lebowitz series, in press.

\bibitem{GoldbergTarjan}
A. V. Goldberg and R. E. Tarjan, J.~ACM {\bf 35}, 921 (1988).

\bibitem{mcnamarathesis}
D. L. McNamara, Ph.D. thesis, Syracuse University (1999).

\bibitem{starlabs} This code currently can be found at
\texttt{http://www.star-lab.com/goldberg/index.html}.

\bibitem{dimacschallenge1}
{\em Network Flows and Matching: First DIMACS Implementation Challenge},
David S. Johnson and Catherine C. McGeoch, eds
(American Mathematical Society, Providence, RI, 1993.)
Also see \texttt{ftp://dimacs.rutgers.edu/}\linebreak\texttt{pub/netflow}.

\bibitem{AAMphasepoly}
A. A. Middleton, cond-mat/0104185.

\end{thebibliography}
\end{document}